\documentclass{article}

\usepackage{amsmath}
    \DeclareMathOperator*{\argmin}{arg\,min}

\usepackage{arxiv}
\usepackage[utf8]{inputenc}
\usepackage[T1]{fontenc}
\usepackage[scale=0.925]{sourcecodepro}
\usepackage{booktabs}
\usepackage{amsfonts}
\usepackage{microtype}
\usepackage{graphicx}
    \graphicspath{ {./images/} }
\usepackage[authoryear, round]{natbib}
\usepackage{tabularx}
\usepackage{longtable}
\usepackage[export]{adjustbox}
\usepackage{fancyvrb}
    \fvset{
      fontsize=\small,
      xleftmargin=2em
    }
\usepackage[dvipsnames]{xcolor}
\usepackage[
  colorlinks=true,
  allcolors=MidnightBlue
]{hyperref}
\usepackage{caption}
\newcommand{\tran}{^\top}
\newcommand{\invtran}{^{-\top}}

\newcommand{\SparseNUTS}{\texttt{SparseNUTS }}
\newcommand{\package}[1]{\texttt{#1}}
\newcommand{\model}[1]{\textit{#1}}
\newcommand{\func}[1]{\texttt{#1}}
\newcommand{\eqnref}[1]{Equation~\eqref{#1}}
\newcommand{\figref}[1]{Figure~\ref{#1}}
\newcommand{\secref}[1]{\ref{#1}}

\usepackage{setspace}
\title{Leveraging sparsity to improve no-U-turn sampling efficiency for hierarchical Bayesian models}

\author{
 Cole C. Monnahan\\
  Resource Ecology and Fisheries Management\\
  Alaska Fisheries Science Center\\
  NOAA Fisheries\\
  Seattle, WA, USA \\
  \texttt{cole.monnahan@noaa.gov} \\
   \And
 Kasper Kristensen\\
  DTU Aqua\\
  Technical University of Denmark\\
  Copenhagen, Denmark \\
  \texttt{kaskr@dtu.edu} \\
  \And
 James T. Thorson\\
 Resource Ecology and Fisheries Management\\
  Alaska Fisheries Science Center\\
  NOAA Fisheries\\
  Seattle, WA, USA \\
  \texttt{james.thorson@noaa.gov} \\
\And
 Bob Carpenter\\
 Center for Computational Mathematics\\
 Flatiron Institute\\
 New York, New York, USA \\
  \texttt{bcarpenter@flatironinstitute.org} \\
}

\begin{document}
\maketitle
\begin{abstract}
\small\renewcommand{\baselinestretch}{1.15}
Analysts routinely use Bayesian hierarchical models to understand natural processes. The no-U-turn sampler (NUTS) is the most widely used algorithm to sample high-dimensional, continuously differentiable models. But NUTS is slowed by high correlations, especially in high dimensions, limiting the complexity of applied analyses. Here we introduce Sparse NUTS (SNUTS), which preconditions (decorrelates and descales) posteriors using a sparse precision matrix ($Q$). We use Template Model Builder (TMB) to efficiently compute $Q$ from the mode of the Laplace approximation to the marginal posterior, then pass the preconditioned posterior to NUTS through the Bayesian software Stan for sampling. We apply SNUTS to seventeen diverse case studies to demonstrate that preconditioning with $Q$ converges one to two orders of magnitude faster than Stan's industry standard diagonal or dense preconditioners. SNUTS also outperforms preconditioning with the inverse of the covariance estimated with Pathfinder variational inference.  SNUTS does not improve sampling efficiency for models with the highly varying curvature found in funnels, wide tails, or multiple modes.  SNUTS is most advantageous, and can be scaled beyond $10^4$ parameters, in the presence of high dimensionality, sparseness, and high correlations, all of which are widespread in applied statistics. An open-source implementation of SNUTS is provided in the R package \package{SparseNUTS}.
\end{abstract}
\renewcommand{\baselinestretch}{1}

\newpage
\section{Introduction}
Hierarchical models are used to study varied processes due to their flexibility in modeling multi-level processes \citep{cressie2009}. Many of them are Bayesian hierarchical analyses which typically use Markov chain Monte Carlo (MCMC) algorithms to generate posterior samples for inference \citep{gelman2014}. The no-U-turn sampler (NUTS; \cite{hoffman2014}) is a widely-used MCMC algorithm for complex hierarchical models (e.g., \cite{monnahan2017}), but its efficiency (effective posterior samples generated per time) can be undermined by a variety of challenging posterior geometries which limits practical applications due to burdensome run times and lack of robust convergence \citep{betancourt2015}. Thus, developing algorithms and strategies to improve NUTS sampling efficiency and reduce run times is important for applied studies in varied fields.

\begin{figure}[ht]
\centering
\includegraphics[scale=.8]{"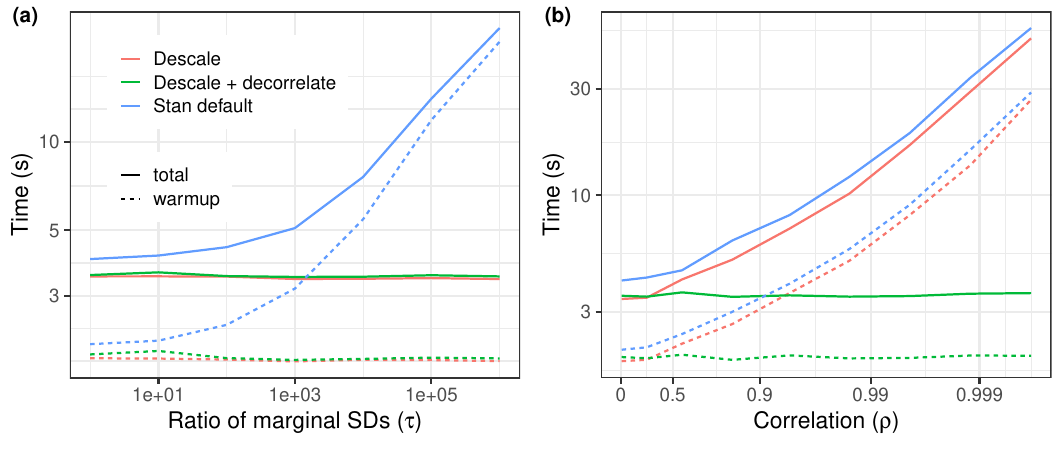"}
\caption{{\bfseries Warmup time comparisons.} Simple illustration of the impact of the ratio of marginal standard deviations (a) and correlations (b) on warmup (dashed lines) and total NUTS run time (solid lines) for a simple bivariate normal model $\mathbf{X} \sim \mathcal{N}(\mathbf{0},\Sigma)$. In (a) $\Sigma=\big(\begin{smallmatrix}
  1 & 0\\
  0 & \tau
\end{smallmatrix}\big)$ and $\tau$ varies, while in (b) $\Sigma=\big(\begin{smallmatrix}
  1 & \rho\\
  \rho & 1
\end{smallmatrix}\big)$ and $\rho$ varies. NUTS was run three times (colors), first using Stan defaults and then after descaling only, or descaling and decorrelating the posterior prior to sampling.  For all cases there were an initial 1000 warmup iterations followed by 1000 sampling iterations and using Stan defaults.  Diagonal mass matrix adaptation was engaged during the warmup period.}
\label{fig:example}
\end{figure}

One challenge is large (linear) correlations and differences in marginal scales among parameters (\figref{fig:example}), which is represented as a condition number factor $\kappa = ( \sum_{d=1}^D ( \frac{\max(\lambda)}{\lambda_d} )^4 )^{1/4}$, where $\lambda$ is the sequence of eigenvalues of the inverse negative Hessian (i.e., local covariance) of the log density evaluated at the current position \citep{langmore2019condition}. Condition factors are a proxy for how difficult NUTS sampling will be, with values close to one easier, and both high correlation and widely varying scales lead to large $\kappa$ and inefficient sampling.\footnote{For example, the eigenvalues of a two-dimensional multivariate normal with unit standard deviation and correlation $\rho$ are $1-\rho$ and $1 + \rho$, the ratio of which is unbounded.  Similarly with zero correlation and scales $\sigma_1, \sigma_2$, the eigenvalues are $\sigma_1^2$ and $\sigma_2^2$, the ratio of which is also unbounded.} The global covariance can be used to precondition a model (i.e., decorrelate and descale) and improve sampling efficiency, but is generally not known prior to sampling and can be prohibitive in higher dimensions due to the costs of matrix algebra \citep{betancourt2017,neal2011,bales2019}. In such cases, full NUTS sampling can be avoided with approximate algorithms like automatic differentiation variational inference (ADVI; \cite{kucukelbir2017}) or Pathfinder \citep{zhang2022}, but these are not widely used because of algorithmic instability and difficulties validating their fits. Another option is to estimate the global covariance during the warmup phase, but this remains a challenge despite low-rank approximations \citep{bales2019}. Yet another approach is to use NUTS to sample from the marginal posterior using the embedded Laplace approximation approach (ELA; e.g., \cite{margossian2020}), which shows promise but has not been widely tested. Consequently, the defaults used by popular platforms like Stan and PyMC is a long warmup NUTS phase to estimate and adjust for parameter scales, but not correlations \citep{carpenter2017, stan2025, abril2023}.  It would thus be extremely useful to have a scalable way to adjust for correlations and scales prior to NUTS sampling.

Recent innovations in maximum marginal likelihood estimation (MMLE) for hierarchical models has increased the size and complexity of models possible. In particular, Template Model Builder (\package{TMB}; \cite{kristensen2016}) is used for popular classes of hierarchical models, such as generalized linear mixed models (GLMMs;  \cite{brooks2017} and spatio-temporal regression and species distribution models \citep{thorson2025b,anderson2025,kellner2023}, as well as niche applications like state-space population dynamics, state-space animal movement, mark-recapture and hidden Markov models, and time-series and phylogenetic structural equation models (Table S1). These \package{TMB} models often are formulated to have multivariate normal random effects with sparse precision (i.e., inverse covariance) matrices, such as Gaussian Markov random fields (GMRF; \cite{rue2005}) and the stochastic partial differential equation (SPDE) approximation widely-used in spatial models \citep{lindgren2011}.  \package{TMB}'s increasing popularity is due to its ability to automatically detect and exploit sparsity for efficient MMLE, but also the calculation of a sparse precision matrix ($Q$) representing the approximate precision of the parameters.  \package{TMB}'s success using $Q$ for inference across diverse hierarchical models suggests that it could be used to account for global correlations and variable scales that challenge applied Bayesian analyses.

Here, we develop the SNUTS algorithm by joining \package{TMB}'s existing computational tools with Stan's NUTS algorithms, while implementing a novel sparse preconditioner to improve NUTS' sampling efficiency.  We benchmark TMB against Pathfinder for generating approximate samples, and then compare SNUTS efficiency against NUTS using Stan defaults for two simulated models of increasing dimensionality up to $10^4$. We then benchmark SNUTS and embedded Laplace approximation SNUTS on seventeen case studies of various complexity, dimensionality, and sparseness. Finally, we implement SNUTS in a new R package \package{SparseNUTS} and demonstrate its use for \package{TMB} and \package{RTMB} \citep{rtmb2025} models.

\section{Methods}\label{sec:methods}
First, we review why parameter correlations undermine NUTS sampling efficiency. Second, we develop a new approach to precondition a posterior with sparse matrix algebra, and review how TMB estimates $Q$. Third, we introduce a new R package \package{SparseNUTS} which provides a user-friendly workflow for Bayesian analysis, including heuristics to automatically select a preconditioner and the length of warmup needed. Finally, we use \package{SparseNUTS} to evaluate how much more efficient SNUTS is that standard NUTS.

\subsection{Sparse preconditioning in NUTS}
\subsubsection{An overview of HMC}
To understand why correlations degrade NUTS sampling efficiency we first review the underlying algorithm. NUTS belongs to the Hamiltonian Monte Carlo (HMC) family of MCMC algorithms which use Hamiltonian dynamics to efficiently generate proposals \citep{neal2011, betancourt2017}.  Hamiltonian dynamics describe the movement of a fictitious puck sliding over a frictionless surface through time governed by the Hamiltonian function $H(p,q)=U(q)+K(p)$.  $K(p)$ is the kinetic energy function on momentum vector $p$ which typically is $K(p) = p\tran M^{-1} p / 2$  where $M$ is a symmetric, positive-definite ``mass matrix.''  $U(q)$ is the potential energy of the position vector $q$. In HMC, $U(q)=-f(q)$ is used where $f(q)$ is the unnormalized log posterior density function and $q$ the parameter vector.

Solutions to the Hamiltonian system are usually approximated with discrete steps using the leapfrog algorithm, which alternates updates of $p$ and $q$  \citep{neal2011}. A single leapfrog step consists of three updates using a step size $\epsilon > 0$ and $M$ starting from position and momenta $(q_t, p_t)$,
\begin{align}\label{eq:leapfrog}
 p_{t+\epsilon/2}  &= p_t + \frac{\epsilon}{2}\nabla f(q) \nonumber \\
 q_{t+\epsilon}  &=q_{t} + \epsilon M^{-1} p_{t + \epsilon/2} \\
 p_{t+\epsilon}  &=p_{t+\epsilon/2} + \frac{\epsilon}{2} \nabla  f(q). \nonumber
\end{align}

A sequence of $L$ leapfrog steps generates an approximate HMC ``trajectory'' through time, from which a point can be selected to generate MCMC samples \citep{betancourt2016}. The NUTS algorithm avoids the need to specify $\epsilon$ because it is adapted, and also $L$ which is determined dynamically by stopping when a trajectory makes a ``U-turn'' \citep{hoffman2014}.  The goal is to make directed movement across the posterior distribution.  Generating points with a high expected squared jump distance directly improves effective sample size by avoiding diffusive random walk behavior, making NUTS efficient at generating posterior samples \citep{neal2011, hoffman2014}. Importantly,  NUTS requires evaluating the gradient function $\nabla f(q)$ to generate trajectories (\eqnref{eq:leapfrog}), which is the computational bottleneck, consuming over 90\% of the compute budget in tight implementations. Thus any reduction in the number of evaluations required for trajectories to move similar distances will improve NUTS performance.

NUTS trajectories are rarely able to span the posterior in the presence of correlations due to oscillatory behavior, caps on the maximum number of leapfrog steps, and premature U-turns.  Often trajectories require many gradient evaluations (Figure S1), which slows sampling. To the extent we can precondition the posterior to make its condition number closer to unity, we can lengthen NUTS trajectories at low cost and greatly improve sampling efficiency.  In the ideal case, a posterior of unit scale and no correlation has a unit condition number.  If prior information about the global covariance of the posterior is known, and it does not vary too widely over the posterior, we can use it to precondition the posterior. Specifically, the parameter space can be linearly transformed to $q'=L^{-1}q$ where $L=\text{chol}(\Sigma)$ is the lower-triangular Cholesky decomposition such that $\text{cov}(q)=\Sigma=L  L\tran$, while $M$ is taken to be the identity matrix, so that the kinetic energy function is $K(p)=p\tran p/2$ \citep{neal2011}. This follows because, abusing random variable notation slightly following \cite{gelman2014}, we have
\[
\text{cov}(q')
\ = \
\text{cov}(L^{-1}  q)
\ = \
L^{-1}  \text{cov}(q)  L\invtran
\ = \
L^{-1}  L  L\tran  L\invtran
\ = \
\textrm{I},
\]
which has condition number one as desired. Preconditioning will thus lead to efficiency gains whenever the posterior curvature varies little as is the case for any distribution that is approximately multivariate normal, because normal distributions have constant curvature.

\subsubsection{Preconditioning with a dense matrix}
In order to use NUTS to sample from $q'$, the log density $f_{q'}(q')$ and gradient $g_{q'}(q')$ functions  are needed to generate trajectories. These can be derived as:
\begin{align}\label{eq:dense}
  f_{q'}(q') &= f_q(L q') \\
  g_{q'}(q') &= \nabla f_{q'}(q')=g_q(L q')  L,
\end{align}
where the constant change-of-variables adjustment factor $\left|\text{det}(L^{-1})\right|$ from the transformation can be dropped (see section 4.1 of \cite{neal2011}). $\Sigma$ will typically be a dense matrix and so $L$ will be dense as well, leading to a computational cost to compute the product $L  q'$ of $\mathcal{O}(D^2)$ for dimension $D$. This quadratic scaling with dimension motivates setting the off-diagonal elements of $\Sigma$ to zero such that $L$ is a diagonal matrix and the matrix-vector product $L  q'$ reduce to element-wise vector product which scales as $\mathcal{O}(D)$. We refer to these two cases of $\Sigma$ as ``dense'' and ``diagonal'', respectively, mirroring the terminology used by Stan \citep{stan2025}. While computationally faster, a diagonal matrix will only account for variable scales and not correlations which can lead to long NUTS trajectories and inefficient sampling (\figref{fig:example}; \cite{neal2011}).

\begin{figure}[ht]
\includegraphics[scale=.9, center]{"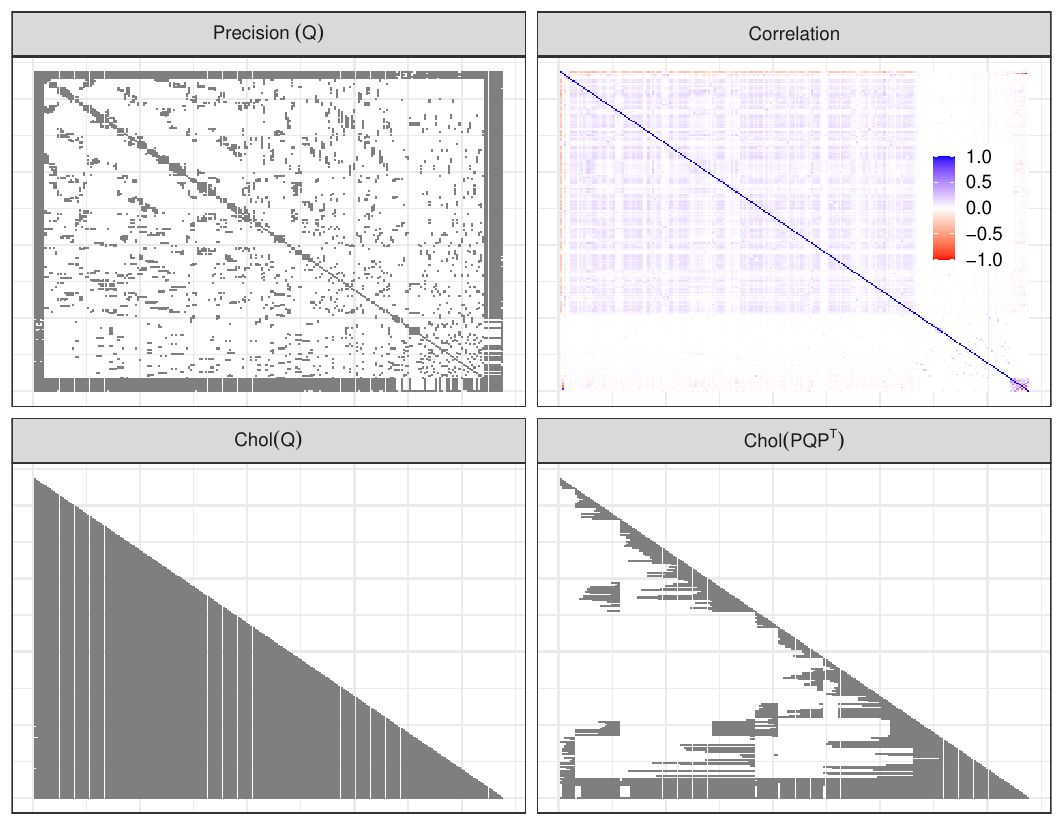"}
\caption{{\bfseries Visualizations of Q.}
The resulting sparsity pattern detected automatically by TMB for the precision matrix $Q$ (top left),  correlation matrix derived from $\Sigma=Q^{-1}$ (top right), the lower triangle Cholesky decomposition of  $Q$ itself (bottom left), and the Cholesky decomposition of $Q$ after permuting by the matrix $P$ to maximize sparsity (i.e., $PQP\tran$, bottom right)  for the \model{sdmTMB} spatial model. Non-zeroes are colored grey while zeroes are white (to highlight sparsity) except the correlation matrix which is colored by the correlation values. $Q$ has 81\% off-diagonal zeroes, while the correlation matrix is fully dense, demonstrating the difference in sparsity between $Q$ and $Q^{-1}$.  In this case the sparsity pattern in $Q$ leads to a nearly dense Cholesky (bottom left), but sparsity is recovered by first permuting $Q$ (bottom right)}
\label{fig:Q_L_cor}
\end{figure}

\subsubsection{Efficient preconditioning with a sparse matrix}\label{sec:smetric}
We can precondition in a similar way but utilizing the sparse nature of $Q$ by letting $q'=L\tran q$ where $L=\text{chol}(Q)$ and $Q=LL\tran$.  Importantly, the amount of sparsity in $L$ will depend on the sparsity of $Q$, but also its ordering because the number of possible non-zero elements of $L$ is greater than the nonzero elements in the lower triangle of $Q$ (see section 2.4 of \cite{rue2005}). The amount of sparsity in $L$ can be increased by reordering $Q$ via a permutation matrix $P$ for which pre-multiplication reorders rows and post-multiplication by the transpose reorders columns, i.e., $Q_P=PQP\tran$. This permutation can dramatically increase sparsity in $L$ and reduce the computational cost of the sparse matrix multiplications below (\cite{rue2005}; \figref{fig:Q_L_cor}).

To implement this sparse preconditioning, we first use the R function \func{Cholesky} from the \package{Matrix} R package \cite{matrix2025} to determine $P$.  Then we let $L=\text{chol}(Q_P)$ so that $Q_P=LL\tran$, and define $q'=L\tran Pq$  and, again ignoring the change of variables constant, define the sparse preconditioner as
\begin{eqnarray}
  f_{q'}(q')&=&f_{q}((L\tran P)^{-1}q') \nonumber\\
  g_{q'}(q')&=&\nabla f_{q'}(q')=
    g_{q}((L\tran P)^{-1} q')(L\tran P)^{-1}.
    \label{eq:perm}
\end{eqnarray}
To calculate $\text{cov}(q')$, first note that $Q_P^{-1} = (L  L\tran)^{-1} = L\invtran \,L^{-1}$, so
\[
\text{cov}(q) = Q^{-1} =(P^{-1}  Q_PP\invtran)^{-1} = P\tran  L  \invtran L^{-1}  P,
\]
Substituting in the preconditioned variable yields
\[
\text{cov}(q')=\text{cov}(L\tran P  q)=L\tran P  Q^{-1}(L\tran  P)\tran =L  \tran  P  (P\tran  L\invtran L^{-1}  P)P\tran  L = \mathrm{I},
\]
because permutation matrices satisfy $P  P\tran=\textrm{I}$. Thus $q'$ has a condition factor of one as desired.  Efficient implementation of these sparse matrix calculations was done in R using the \func{solve} function from the \package{Matrix} package, with example code in Appendix C.  %Importantly, the product $L^{\invtran}q'$ can be calculated without explicitly calculating $L^{\invtran}$ through inversion using efficient sparse matrix algorithms such as forward substitution \citep{scott2023}.

\begin{figure}[ht]
\includegraphics[scale=.9, center]{"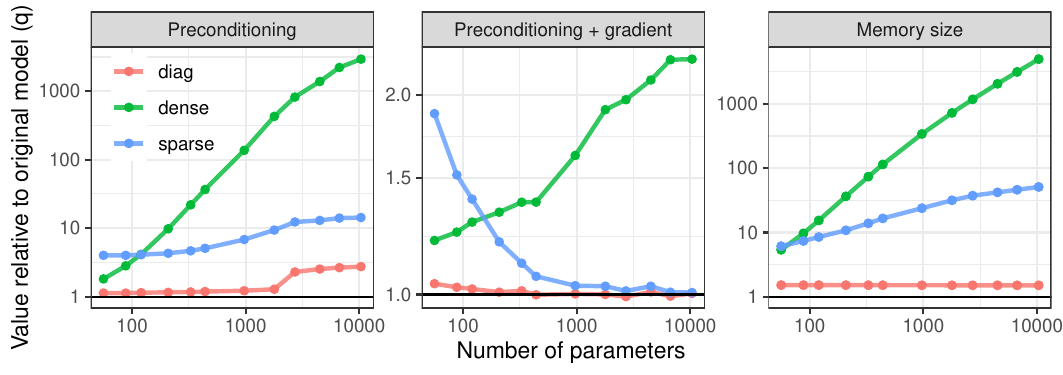"}
\caption{{\bfseries Benchmark results.}  Benchmark results for a simulated Poisson SPDE model of increasing number of observations and spatial random effects with high sparsity of $Q$ by design, typically > 99\%. The plots show: left) the computational cost of just the transformation of $g_q \rightarrow g_{q'}$ alone (i.e., with a negligible cost of evaluating $g_q$); center) the transformation and the cost of the gradient $g_q(q)$; and right) the memory used for the transformation. Three types of transformations (colors) are shown relative to the value from the original model.}
\label{fig:gr_bench}
\end{figure}

The computational and memory costs of \eqnref{eq:perm} will depend on the dimensionality and amount of sparsity in $(L\tran P)^{-1}$, so we tested this by benchmarking the gradient evaluations outside of NUTS for a diagonal, dense, and sparse matrices on a spatial SPDE model (which has high sparsity in $Q$), with increasing dimension.  The sparse decorrelation was slower than a dense one in lower dimensions, but faster for models with $d>100$ (\figref{fig:gr_bench}), and overall that the computational cost of $g_{q'}(q')$ was low compared to  $g_q(q)$, meaning the sparse preconditioning adds negligible computational cost for $d>10^4$, at least for this particular model. In \secref{sec:sims} we evaluate the performance on NUTS sampling efficiency in more detail.

We now turn to how a sparse precision matrix $Q$ can be found prior to NUTS sampling so that the sparse preconditioning can be tested against the dense and diagonal versions. \package{TMB} is, to our knowledge, the only software platform that can calculate $Q$ with sparsity for arbitrary statistical models and so we focus on it here. \package{TMB} itself, and the frequentist techniques it uses, may be unfamiliar to some readers so we provide a brief overview, and then detail how $Q$ is calculated and used in a Bayesian context.

\subsubsection{Template Model Builder overview} \label{sec:tmb}
Template Model Builder (\package{TMB}) is a statistical software package that performs MMLE for high dimensional, non-linear hierarchical models \citep{kristensen2016}. Users write negative log-likelihoods in templated C++ which is then compiled and linked via an R interface. Alternatively the user can write models in R directly via the \package{RTMB} interface to \package{TMB} \citep{rtmb2025}. The user also declares which parameters are ``random effects'' (i.e., latent states which are typically assumed to have Gaussian priors) such that the parameter space $x=(u,\theta)$ is partitioned into random ($u$) and fixed ($\theta$) effects, and $f_x(u,\theta)$ is the negative log-likelihood of the data and random effects, and $g_x(u,\theta)=\nabla f_x(u,\theta)$ is the gradient function.

The MMLE of the fixed effects  $\hat\theta$ are found by maximizing the Laplace approximation to the marginal likelihood $L^*(\theta)=\sqrt{2\pi}^n \det(H(\theta))^{-1/2}\exp(-f_x(\hat{u},\theta))$ \citep{skaug2006}, where  $\hat{u}(\theta)=\argmin_u f_x(u,\theta)$ is the $u$ that minimizes the objective function while holding $\theta$ fixed, and $H(\theta)=f^{''}_{uu}(\hat{u}(\theta), \theta)$ is the Hessian of $f_x(u,\theta)$ w.r.t. $u$ and evaluated at $\hat{u}(\theta)$. \package{TMB}  evaluates $f_\theta(\theta)=-\log L^*(\theta)$ and its gradient $g_\theta(\theta)$ with automatic differentiation and utilizing sparsity of $u$.  $\hat{\theta}$  is referred to as the marginal mode, $\hat{u}(\hat{\theta})$ the empirical Bayes estimates for $u$, and  $\hat{x}=(\hat{u}(\hat{\theta}),\hat{\theta})$ the conditional mode.

\subsubsection{Calculating a sparse approximate precision matrix}\label{sec:estQ}
For notational simplicity we assume $u$ is the first component of the parameter vector. Then, $Q$ is the matrix of second derivatives evaluated at $\hat{x}$ represented in block form as:

\begin{equation}
 Q=\begin{bmatrix}
H_{uu} & H_{u\theta} \\
 H_{\theta u} & H_{\theta\theta}
\end{bmatrix}
\label{eq:estQ}
\end{equation}

where $H_{uu}$ is the sparse Hessian of $u$ at $\hat x$,  $H_{u\theta}$ and $H_{\theta u}=H_{u\theta}\tran$ the fixed and random effect block of the Hessian evaluated at  $\hat{x}$. Finally,  $H_{\theta\theta}=H_{\theta u} H_{uu}^{-1} H_{u \theta} + H_{\hat{\theta}\hat{\theta}}$ where $H_{\hat\theta\hat\theta}$ is the Hessian of the marginal likelihood $f_\theta$ at the marginal mode $\hat\theta$ \citep{kass1989}. \package{TMB} exploits sparsity and automatic differentiation throughout, and the result is a sparse matrix type from the \package{Matrix} package.

An element  $Q_{ij}=0$ when the $i^\text{th}$ and $j^{\text{th}}$ parameters are independent conditional upon fixed values for all other parameters, detected automatically from a graph representing model computations using the ``subgraph'' algorithm \citep{bell2021}.  The sparsity pattern of $Q$ thus reflects the conditional independence structure of the model (e.g., Figure \ref{fig:Q_L_cor}). In the frequentist framework $Q$ provides estimates of mean squared errors based on a multivariate normal approximation of the distribution of $u$ given the data, rather than variance estimates due to repeated sampling of the data and $u$ \citep{zheng2021}.  In the Bayesian context we assume $Q$ is an approximation to the posterior precision.

\subsection{The sparse NUTS algorithm}\label{sec:snuts}
\package{TMB} can be used in a Bayesian context by incorporating priors and Jacobian corrections for parameter transformations into the objective function. This converts the unnormalized negative log-likelihood to the unnormalized log-posterior, $f_q(q)=-f_x(u,\theta)$ and $g_q(q)=-g_x(u,\theta)$, while maintaining the fixed ($\theta$) and random ($u$) effect partitioning of the Bayesian parameter vector $q=(u,\theta)$. $\hat{q}=(\hat{u(\theta)}, \hat{\theta})$ is thus the conditional posterior mode (as defined above), and $Q$ is assumed to approximate the posterior as $q\sim \mathcal{N}(\hat{q},Q^{-1})$. Likewise, the generalized delta method can be used to get approximate posterior distributions for functions of parameters \citep{kristensen2016}, with a worked example in Appendix C.

We propose the following steps for SNUTS:
\begin{enumerate}
\item Write out the unnormalized log-posterior density $f_q(q)$, specifying the random effects $u$ and thus partitioning $q=(u,\theta)$.
\item  Numerically maximize the Laplace approximation of the marginal posterior and calculate $Q$ (see \eqnref{eq:estQ}) for hierarchical models, or $\Sigma$ for models without random effects, at the mode $\hat{q}=(\hat{u}(\hat{\theta}), \hat{\theta})$.
\item Use $Q$ and $\Sigma$ to automatically select the preconditioner and length of warmup, and  get $f_{q'}$ and $g_{q'}$ via \eqnref{eq:dense} or \eqnref{eq:perm}.
  \item	Pass $f_{q'}$ and $g_{q'}$ to a NUTS algorithm to generate posterior samples of $q'$. Here we use Stan's NUTS algorithm via the \package{StanEstimators} interface  \citep{johnson2025}.
  \item	Re-condition the parameter space ($q'\rightarrow q$) and assess MCMC convergence via diagnostics and use posterior samples for inference of parameters and derived quantities as in a standard Bayesian analysis.
 \end{enumerate}

We implemented SNUTS in the \texttt{sample\_snuts} function in the open-source R package \SparseNUTS which is available from XXXX
%\href{https://github.com/noaa-afsc/SparseNUTS}{https://github.com/noaa-afsc/SparseNUTS}
and provides a user-friendly interface, including automatic selection of the preconditioner and length of warmup (see Appendix A for more details and a worked example).

\subsection{Evaluations}\label{sec:evaluations}
\subsubsection{Case studies}
We evaluate SNUTS on 17 models with different structures, levels of data, and dimensions (Table~\ref{tab:mods}).  Table~S2 provides further details, any modifications, and original citations. %Code and model files to recreate this study can be found online at \url{https://github.com/Cole-Monnahan-NOAA/sparse_nuts}.

%\begin{landscape}
\begin{table}[ht]
\footnotesize
\centering
\begin{tabular}{p{1.5cm}p{4.15cm}p{1.5cm}p{1.0cm}p{2.2cm}p{2cm}p{1cm}}%{\textwidth}{|l|l|X|X|X|X|X|}
\hline
\textbf{Name} & \textbf{Model type} & \textbf{No. pars.} & \textbf{Sparsity} & \textbf{Max correlation} & \textbf{SD ratio} & \textbf{Preconditioner} \\
\hline
% paste table into this website then copy only the content rows https://www.tablesgenerator.com/
causal         & Causal time series             & 264 (232)    & 78.7    & 0.96, 0.97        & 101.8, 111.0   & sparse \\
diamonds       & Linear mixed model              & 26 (24)     & 0    & 0.89, 0.89           & 189.6, 189.7   & dense  \\
dlm            & Dynamic linear model            & 118 (114)   & 91.7 & 0.67, 0.67           & 9.2, 9.5       & dense  \\
gp\_pois\_regr & Poisson Gaussian process        & 13 (11)     & 0    & 0.972, 0.988         & 8.1, 8.4       & dense  \\
irt\_2pl       & Item response theory            & 148 (144)   & 56.3 & 0.84, 0.68           & 8.4, 13.2      & dense  \\
kilpisjarvi    & Linear regression               & 3 (0)       & 0    & 0.9999887, 0.9999884 & 3982.5, 3982.5 & dense  \\
lynx           & Runge-Kutta  ODE solver                  & 50 (42)     & 63   & 0.89, 0.88           & 19.5, 19.2     & dense  \\
petrel         & Hidden Markov model             & 274 (266)   & 94.2 & 0.97, 0.97           & 115.8, 114.3   & dense  \\
pollock        & Non-linear population dynamics  & 351 (55)    & 28.1 & 0.9999821, 0.9999817 & 66.3, 65.2     & sparse \\
radon          & Linear mixed model              & 389 (386)   & 98.5 & 0.27, 0.12           & 79.4, 75.7     & diag   \\
salamanders    & Zero-inflated negative binomial            & 39 (23)     & 44.5 & 0.963, 0.958         & 7.6, 6.9       & dense  \\
sam            & Non-linear population dynamics  & 1226 (1209) & 96.6 & 0.91, 0.91           & 11.9, 12.0     & sparse \\
schools        & Linear mixed model              & 10 (8)      & 66.2 & 0.26, 0.29           & 6.1, 6.45      & diag   \\
sdmTMB         & Spatial Tweedie GLMM            & 219 (216)   & 81.3 & 0.94, 0.92           & 451.0, 352.2   & dense  \\
swallows       & State-space mark-recapture      & 177 (148)   & 75.4 & 0.94, 0.96           & 37.3, 37.5     & dense  \\
wham           & Non-linear population dynamics  & 385 (322)   & 66.5 & 0.93, 0.91           & 26.5, 9.9      & sparse \\
wildf          & Binomial GLMM                   & 1101 (1092) & 98.4 & 0.96, 0.95           & 15.6, 11.8     & sparse \\[8pt]
\hline
\end{tabular}
\vspace*{4pt}
\caption{{\bfseries Case studies}. No. parameters given is total and those considered random effects in parentheses. Sparsity in $Q$ is the percentage of known zero elements in the lower triangle of $Q$ excluding the diagonal. The max correlation is the maximum of absolute pairwise correlations among all parameters for the posterior followed by that estimated from $Q$. The SD ratio is the ratio of marginal parameter standard deviations from the posterior and from $Q$. The preconditioner column shows the automatically selected approach.}
\label{tab:mods}
\end{table}
%\end{landscape}

\subsubsection[Quantifying the accuracy of Q]{Quantifying the accuracy of $Q$}
A key claim is that $\mathcal{N}(\hat{q},Q)$ approximates the posterior geometry well for many hierarchical models. We test this claim by evaluating it against Pathfinder, which efficiently generates approximate samples from the posterior distribution and was developed primarily to initialize NUTS chains \citep{zhang2022}. This approach was found to outperform ADVI algorithms across a variety of models and so we use it as a reference method \citep{zhang2022}. We compare Pathfinder's performance against TMB's approach of multivariate normal samples $q^{sim}\sim \mathcal{N}(\hat{q},Q^{-1})$ which we refer to as the ``precision sampling'' approach. We gauge accuracy of these two methods using the 1-Wasserstein distance \citep{mccann1995,craig2016} where larger values indicate a worse posterior approximation. We also compare computational costs (run time), with specific details in Appendix A.

\subsubsection{Comparing sampling efficiency between algorithms}
Sampling efficiency for an MCMC algorithm is the time required to generate an effective posterior sample, and is a common metric to compare performance among different algorithms or software platforms (e.g., \cite{monnahan2017}). Effective sample size (ESS) is defined as the minimum across all model parameters and the log-posterior density, and was calculated from the \func{ess\_bulk} function from the \package{posterior} R package \citep{buerkner2025} which implements the method in  \cite{vehtari2021}. Run time excludes any model compilation, and is the sum of warmup and sampling time across parallel chains, as well as the time to optimize, calculate and invert $Q$ as part of the automatic selection algorithm (i.e., SNUTS overhead).

To benchmark SNUTS against NUTS in the same framework below (i.e., \func{sample\_snuts}) we ran each analysis twice, once with Stan defaults which is 1000 warmup iterations during which a diagonal mass matrix is adapted and 1000 sampling iterations, which we refer to as ``NUTS'' subsequently. And then again with the SNUTS algorithm with automatic selection of the preconditioner (i.e., diagonal, dense or sparse) , 150 warmup samples, disabled mass matrix adaptation, and 1000 sampling iterations. All chains were initialized from random draws from $\mathcal{N}(\hat{q},Q^{-1})$. We ran three independent analyses of four chains to quantify the variance in results due to the different initial values and stochastic nature of NUTS.

\subsubsection{Performance in higher dimensions}\label{sec:sims}
We evaluated the performance of SNUTS with increasing dimensionality on two different models. First, we replicated the analysis from \cite{brooks2017} where data were simulated from a negative binomial GLMM (see \model{salamanders} in Table~\ref{tab:mods}) with increasing number of sites and thus random effects. Second, we simulated from a Poisson spatial GLMM parameterized as an SPDE model, as benchmarked above in \figref{fig:gr_bench}, with increasing observations and random effects.

\subsubsection{Benchmarking sparse NUTS performance}
Next we evaluated the performance of SNUTS on our case studies (Table~\ref{tab:mods}). First, we quantified the overhead of SNUTS by comparing gradient costs with and without preconditioning as well as the time to estimate $Q$ relative to a full SNUTS run. We then ran NUTS and SNUTS on each case study, again using three analyses of four chains.

\subsubsection{Embedded Laplace approximation}
Using NUTS to sample from the marginal posterior ($f_\theta$) was first explored in \cite{monnahan2018} and in then in more depth by \cite{margossian2020} who called this approach ``embedded Laplace approximation'' (ELA) HMC. The ELA approach is trivial to implement in \SparseNUTS by setting the \texttt{sample\_snuts(..., laplace=TRUE)}.  We excluded models \model{petrel}, \model{pollock}, and \model{wham} which were deemed too slow. We compared sampling efficiency of ELA-NUTS and ELA-SNUTS against full SNUTS.  Differences in posterior distributions of fixed effects of ELA NUTS against NUTS can expose bias due to the inaccuracy of the Laplace approximation \citep{monnahan2018}. Our focus is on computational efficiency of NUTS, so we did not investigate bias, but we note it is an important consideration of using ELA-NUTS.

\section{Results}\label{sec:results}
\begin{figure}[ht]
\includegraphics[scale=.9, center]{"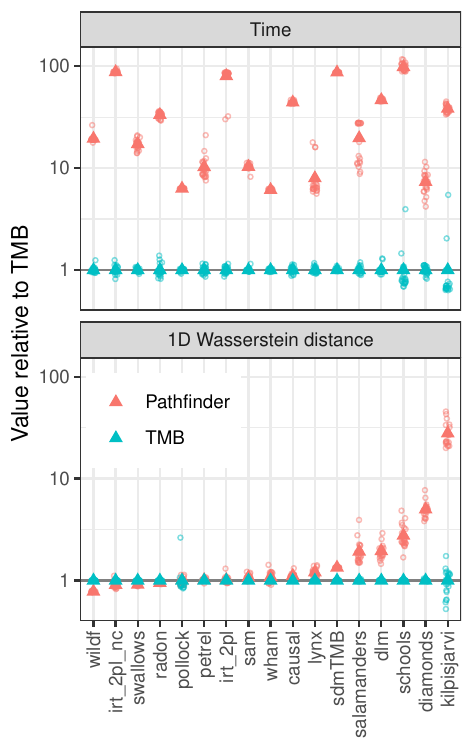"}
\caption{{\bfseries Comparison to Pathfinder.} The plots compare the approximation performance of \package{TMB}'s approximation against Pathfinder for the case studies given in Table~\ref{tab:mods}. The \package{TMB} method assumes the posterior is approximated by $\mathcal{N}(\hat{x}, Q^{-1})$ where $\hat{x}$ is the conditional mode and $Q$ the sparse precision matrix calculated in equation \ref{eq:estQ}.  Each of the 20 points represents a single analysis with a different random initial parameter vector (a posterior sample) and random seed for sample generation. Means are given by filled triangles and is always 1 for \package{TMB}'s method because performance is measured relative to it. Pathfinder used 4 serial paths, while $Q$ optimized once and generated samples. The wall time is shown for the two methods (top; lower is better), and the 1D Wasserstein distance (bottom; lower is better) is a measure of how well the 1000 samples approximate the posterior, with both shown relative to \package{TMB}. Models are ordered by Pathfinder performance relative to \package{TMB}.
}
\label{fig:pfall}
\end{figure}

A comparison of the maximum correlation and ratio of marginal SDs of the posterior indicate that $Q$ was able to accurately capture this aspect of geometry except the \model{wildf} and \model{irt\_2pl} models (Table~\ref{tab:mods}, Figure S4). We found that precision sampling was less sensitive to initial values and substantially faster than Pathfinder for all models (\figref{fig:pfall}a), and provided equivalent or better (i.e., smaller) Wasserstein distances in most models (\figref{fig:pfall}b). These results suggests that sampling from $\mathcal{N}(\hat{q},Q^{-1})$ is both faster, and often better, than those drawn from Pathfinder which in turn was better than other ADVI models \citep{zhang2022}.  $Q$ is thus a promising approximation to a variety of complex, hierarchical posteriors and these samples can be used to initialize SNUTS chains as \texttt{init=``random''} in \func{sample\_snuts} . Since $Q$ must be calculated for SNUTS anyway, these random draws are essentially free.

\begin{figure}[ht]
\includegraphics[scale=.8, center]{"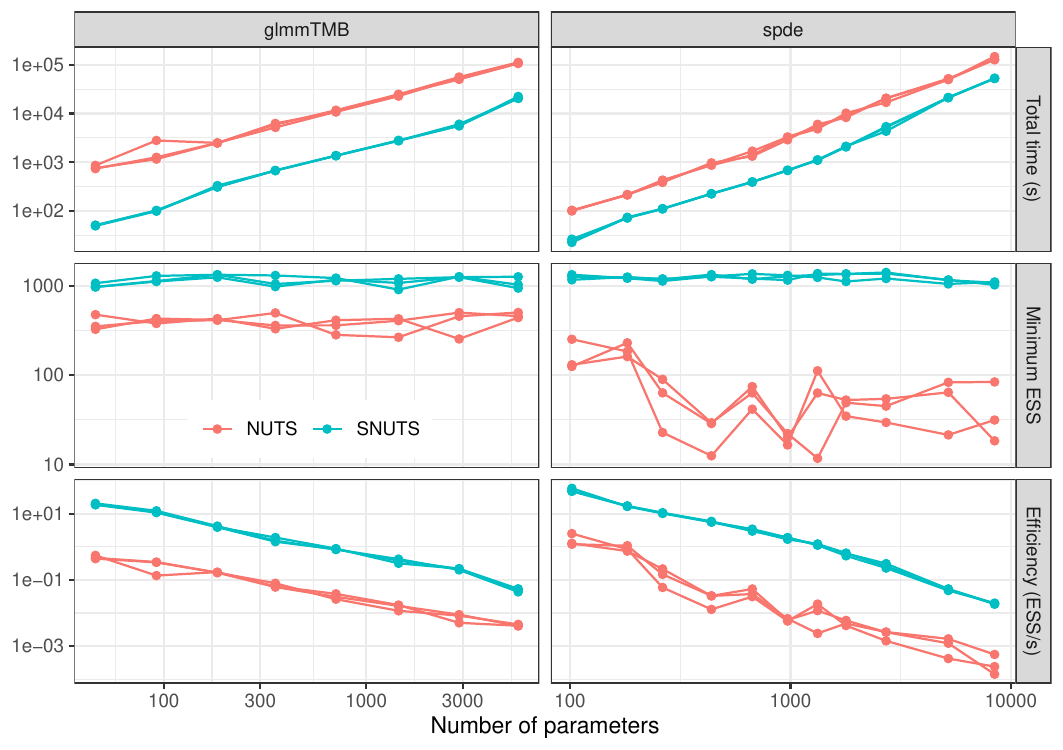"}
\caption{{\bfseries Evaluation of scalability.} Performance results for increasing dimensionality for two models (columns): \package{glmmTMB} is a negative binomial regression and \package{spde} is a spatial Poisson SPDE regression. Different performance metrics (rows) are shown for each algorithm (colors). Three analyses of four chains each (lines) show the variability in estimates. }
\label{fig:perf_sim}
\end{figure}

We found  tests using two simulated examples, a sparse preconditioner was selected in SNUTS except for the \model{glmmTMB} models with fewer than 100 parameters where a dense one was selected. SNUTS took about 6-13\% of the time for the \model{glmmTMB} model with about twice the ESS. Likewise, for the \model{spde} model SNUTS was about 30-70\% of the time and produced 15-20 times the ESS as NUTS. Taken together these faster run times and higher ESS lead to a 20-30 (\model{glmmTMB}) and 20-70 (\model{spde}) fold improvement in SNUTS efficiency over NUTS, depending on the dimensionality (\figref{fig:perf_sim}).

SNUTS also generally outperformed NUTS on our case studies. We found an average computational cost of calculating the transformed gradient $g_{q'}(q')$  ranged from 1.00 (\model{diamonds}) to 1.43 (\model{sam}) times the cost of $g_q(q)$, with most under 1.3 (Table S3). For many of these models the gradient cost is driven largely the cost of evaluating the original gradient $g_q(q)$ which is function of model complexity and amount of data. The overhead for SNUTS was also relatively small,  generally <2\% of the total SNUTS run time for a model. This implies that $Q$ is available prior to NUTS sampling at little added cost.

\begin{figure}[ht]
\includegraphics[scale=.9, center]{"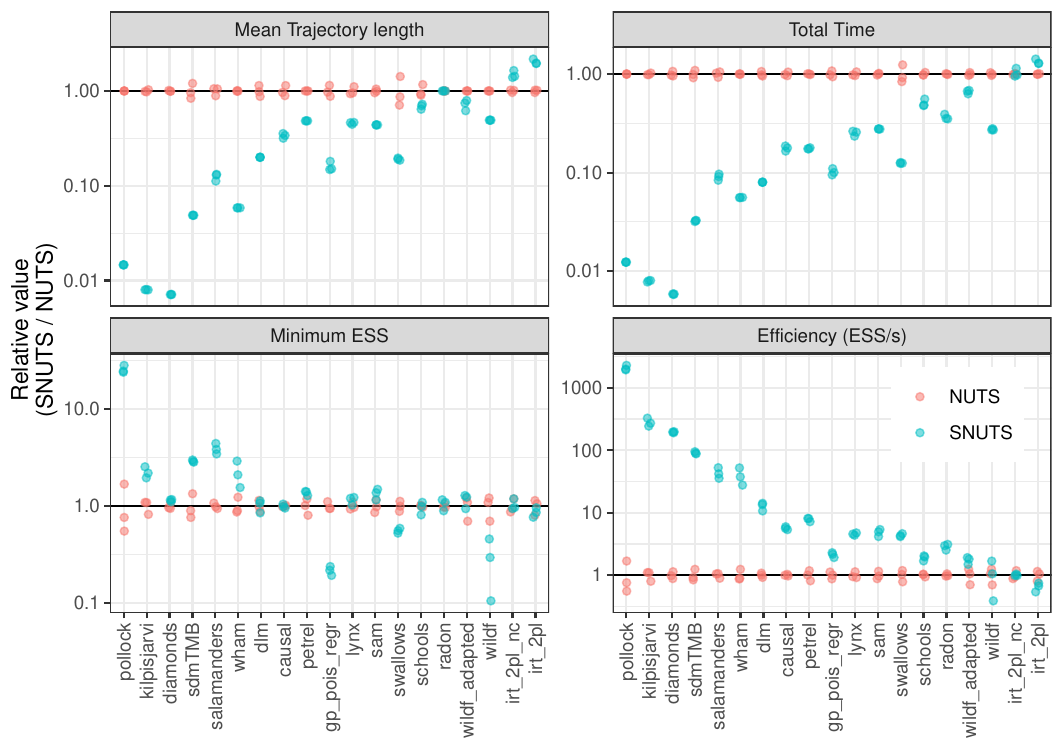"}
\caption{{\bfseries Comparisons to NUTS.} Comparisons of SNUTS to the NUTS default implementation in Stan for four performance metrics over several model/dataset combinations.  Mean trajectory length is the average number of leapfrog steps for post-warmup iterations. Total time (left) includes optimization and calculation of $Q$, sampling and warmup time, but not compilation. The minimum effective sample size (ESS) is the minimum across all parameters. The sampling efficiency is the ratio of minimum ESS to total time. The automatically selected preconditioner for SNUTS (section A1.2) is given in Table~\ref{tab:mods}. Three separate analyses of four chains each (points) show the variation.}
\label{fig:perf_mods}
\end{figure}

We found that SNUTS  had lower total time across most case studies, often by several orders of magnitude, but ESS was more variable between the two algorithms (\figref{fig:perf_mods}). Overall the efficiency (ESS/t) was higher for all but the \model{wildf} and \model{irt\_2pl} models, with performance improvements up to several orders of magnitude with SNUTS. For instance the \model{diamonds} model is listed in \cite{magnusson2024} as having a difficult geometry for NUTS, but SNUTS was 195 times more efficient due to a reduction of average post-warmup trajectory lengths from 979 with NUTS to 7 with SNUTS.  The cost savings of fewer gradient calls dwarfs the SNUTS overhead and slower gradient evaluations for this model. This model has relative few parameters (26) and no sparsity in $Q$, but a similar finding occurs for larger, sparse models. For instance the \model{wham} model, a complicated non-linear state-space population dynamics model, was 39 times more efficient due to a reduction in average trajectory length from 256 to 15 while increasing ESS by 2.19 times.

The \model{wildf} model was notable because using SNUTS resulted in the largest drop in ESS among all models, and also compared the worst to Pathfinder (Figures~\ref{fig:pfall}a, \ref{fig:perf_mods}). In this case it was apparent that $Q$ poorly approximated the posterior due to non-linear correlations and inaccurate marginal variances (parameter scales) so that the preconditioning actually made sampling less efficient (Figure S5).  We turned on adaptation of the diagonal mass matrix in Stan in addition to the sparse preconditioner, allowing Stan to further tune the scales, and indeed this version of the model (\model{wildf\_adapted} in \figref{fig:perf_mods}) was about 1.17 times faster than NUTS.

\begin{figure}[ht]
\includegraphics[scale=0.9, center]{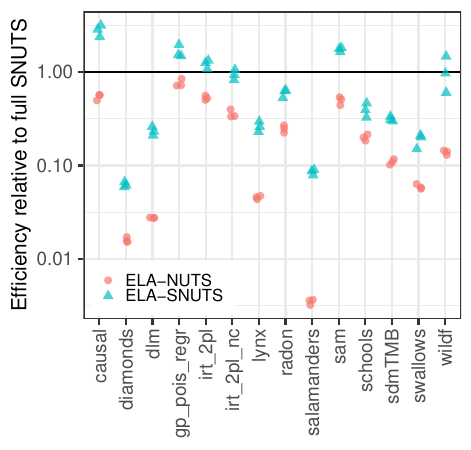}
\caption{{\bfseries Embedded Laplace approximation (ELA) performance.}  ELA is where NUTS or SNUTS samples the posterior marginalized by the Laplace approximation. ELA-NUTS still uses Stan defaults, while ELA-SNUTS preconditions the marginal posterior using an automatically chosen ``dense'' or ``diag'' approach and uses a shorter warmup period without mass matrix adaptation as before (see section \ref{sec:snuts}).  Efficiency (ESS/t) for these two versions of ELA is shown relative to full SNUTS sampling to gauge potential improvements to full SNUTS by using ELA for either NUTS or SNUTS. }
\label{fig:perf_ela}
\end{figure}

Finally, in our testis of  the ELA  approach we found mixed results.  ELA-SNUTS was always more efficient than ELA-NUTS, ranging from 1.97 to 24.4 for the \model{schools} and \model{salamanders} models, respectively (Figure \ref{fig:perf_ela}), suggesting preconditioning the marginal posterior when using ELA is worthwhile. Additionally, full SNUTS outperformed ELA-NUTS for most but not all models; for instance the \model{causal} and \model{sam} ELA models both had a dense preconditioner selected and were 2.79 and 1.74 times more efficient than full SNUTS, respectively. Overall our results suggest that SNUTS will generally outperform ELA-NUTS, and in only some cases will ELA-SNUTS be a better option.

\section{Discussion}\label{sec:discussion}
Conditional independence arises naturally in various hierarchical models and this translates to sparsity in the random effect structures. Packages such as \package{TMB} and \package{lme4} exploit this sparsity to efficiently provide maximum marginal likelihood estimates (MMLE) for a range of hierarchical model forms.  Packages such as \package{INLA} \citep{lindgren2015} have used the same nested Laplace approximation approach for two-stage Bayesian inference.  Here, we took a different approach that uses a global approximation to the entire posterior calculated at the MMLE. Using a Laplace approximation to a Bayesian posterior is not new  (see e.g., \cite{kass1989} and Chapter~13 of \cite{gelman2014}), but preconditioning with a sparse precision matrix calculated at the MMLE is.  Here, we show that TMB can be faster and more accurate than variational inference techniques like Pathfinder (see \figref{fig:pfall}). This accuracy motivates using the posterior approximation in exploratory stages of Bayesian workflow \citep{gelman2020}, or even for inference when high throughput is required.

We focus on using our posterior approximation to globally precondition Hamiltonian Monte Carlo samplers such as the no-U-turn sampler (NUTS).  Here, we show that preconditioning is so effective that we can scale full Bayesian sampling over geometrically challenging hierarchical models to tens of thousands of random effects, which is far beyond what standard inference packages like \package{Stan} or \package{PyMC} can manage with diagonal or dense preconditioners when faced with high correlations. Our approach is black-box in the sense of not needing any customization for particular hierarchical model structures, making it much more widely applicable than a system such as \package{INLA}, which requires custom quadrature rules for each model form.

The sparse precision matrix $Q$ at the heart of our computations is relatively inexpensive to calculate, accounting for less than 2\% of the total SNUTS run time (see Table~S3).  This allows us to account for global correlations in high dimensions with minimal additional overhead (see Figures~\ref{fig:gr_bench}, \ref{fig:perf_sim}). SNUTS samples more efficiently, but that is only part of the story.  Perhaps more importantly, it eliminates the long and brittle warmup periods needed for mass matrix adaptation as is standard in fitting models with NUTS. The bottom line is an improvement in speed of one to two orders of magnitude over the current state of the art for black-box hierarchical models, as well as an improvement in robustness (see \figref{fig:perf_mods}). These improvements are largely possible by exploiting sparsity, and suggests leveraging sparsity can play an important role in many computational aspects of Bayesian inference.

Both sparsity in $Q$ and global correlations were common in the models examined here (Tables \ref{tab:mods}, S1). Whenever correlations are the primary cause of NUTS inefficiency, and that is more often the case than not in hierarchical models that are not data starved, we expect SNUTS to provide robustness, efficiency, and scalability.  This is particularly true for models GMRFs, such as spatial SPDE models, which typically exhibit high correlations and high sparsity (see, e.g., \figref{fig:Q_L_cor}). \package{TMB} serves as the computational backend for popular statistical packages, notably GLMMs, time series models of all kinds, and species distribution models (see Table~S1 for a partial list of such packages).  The existing integration of \package{TMB} means that SNUTS is readily available for a wide class of models. \package{TMB}'s popularity is growing, and with it, the scope of models applicable to SNUTS will continue to grow.

Strong correlations are only one of many causes of inefficient NUTS sampling. Models with performance limited by funnel-like geometry, multiple modes, extreme tails, or other challenging geometries,  will not be well approximated by a multivariate normal distribution and thus SNUTS will likely fail to improve efficiency.  Indeed, the item-response theory two-parameter logistic (\model{irt\_2pl}) model has a notoriously difficult geometry involving multiplicative non-identifiability and funnels due to low data sizes per student, in addition to high correlations. SNUTS was unable to resolve this, even with a non-centered version of the model \cite{papaspiliopoulos2007} (see irt\_2pl\_nc in \figref{fig:perf_mods}).  However, our results here and the asymptotic behavior of Bayesian posteriors as sample sizes increase (see section 4.2 of \cite{gelman2014}) suggest many Bayesian models will be sufficiently normal for SNUTS to improve sampling efficiency, or at least help analysts more quickly identify and resolve other issues caused by non-normal geometries. SNUTS also allows easy implementation of ELA-SNUTS, which could be a useful tool for more challenging geometries.

We implemented SNUTS in \package{TMB} because it automatically detects sparsity for arbitrary models allowing sparse precision matrices $Q$ to be calculated in high dimensions (\eqnref{eq:estQ}). Other platforms utilize sparsity in similar contexts. For instance, \package{INLA} \citep{lindgren2015} does approximate Bayesian inference using a nested Laplace approximation and implements sparse precision matrices to improve computational speed, but is limited to certain classes of latent Gaussian models compared to \package{TMB} (e.g., \cite{osgood2023}). More flexible platforms like Stan and PyMC could benefit from adoption of \package{TMB}'s automatic sparsity detection and Laplace machinery underpinning SNUTS. For now, SNUTS is available for models which \package{TMB} or \package{RTMB} can fit, which includes popular model types, and \package{TMB} can be used to gauge the benefits of adopting sparsity in other Bayesian statistical software packages.

\section{Acknowledgments}
We thank Andrew Johnson for modifying the \package{StanEstimators} package specifically to work with \package{TMB} and \package{RTMB} models. We thanks Th\'eo Michelot for supplying the petrel example. We also thank Darcy Webber and Eric Ward for helpful comments on an earlier version of the manuscript.

\bibliographystyle{unsrtnat}
\bibliography{references}

\begin{thebibliography}{62}
\providecommand{\natexlab}[1]{#1}
\providecommand{\url}[1]{\texttt{#1}}
\expandafter\ifx\csname urlstyle\endcsname\relax
  \providecommand{\doi}[1]{doi: #1}\else
  \providecommand{\doi}{doi: \begingroup \urlstyle{rm}\Url}\fi

\bibitem[Cressie et~al.(2009)Cressie, Calder, Clark, Hoef, and Wikle]{cressie2009}
Noel Cressie, Catherine~A. Calder, James~S. Clark, Jay M.~Ver Hoef, and Christopher~K. Wikle.
\newblock {Accounting for uncertainty in ecological analysis: the strengths and limitations of hierarchical statistical modeling}.
\newblock \emph{Ecological Applications}, 19\penalty0 (3):\penalty0 553--70, 2009.
\newblock \doi{10.1890/07-0744.1}.

\bibitem[Gelman et~al.(2014)Gelman, Carlin, Stern, Dunson, Vehtari, and Rubin]{gelman2014}
Andrew Gelman, John~B. Carlin, Hal~S. Stern, David~B. Dunson, Aki Vehtari, and Donald~B. Rubin.
\newblock \emph{{Bayesian data analysis}}, volume~3.
\newblock Taylor and Francis, 2014.

\bibitem[Hoffman and Gelman(2014)]{hoffman2014}
Matthew~D. Hoffman and Andrew Gelman.
\newblock {{The no-U-turn sampler: adaptively setting path lengths in Hamiltonian Monte Carlo}}.
\newblock \emph{Journal of Machine Learning Research}, 15\penalty0 (1):\penalty0 1593--1623, 2014.

\bibitem[Monnahan et~al.(2017)Monnahan, Thorson, and Branch]{monnahan2017}
Cole~C. Monnahan, James~T. Thorson, and Trevor~A. Branch.
\newblock {{Faster estimation of Bayesian models in ecology using Hamiltonian Monte Carlo}}.
\newblock \emph{Methods in Ecology and Evolution}, 8\penalty0 (3):\penalty0 339--348, 2017.
\newblock \doi{10.1111/2041-210x.12681}.

\bibitem[Betancourt and Girolami(2015)]{betancourt2015}
Michael Betancourt and Mark Girolami.
\newblock {Hamiltonian Monte Carlo for hierarchical models}.
\newblock \emph{Current trends in Bayesian methodology with applications}, 79\penalty0 (30):\penalty0 2--4, 2015.

\bibitem[Langmore et~al.(2019)Langmore, Dikovsky, Geraedts, Norgaard, and Behren]{langmore2019condition}
Ian Langmore, Michael Dikovsky, Scott Geraedts, Peter Norgaard, and Rob~Von Behren.
\newblock {A condition number for {H}amiltonian {M}onte {C}arlo}.
\newblock \emph{arXiv preprint arXiv:1905.09813}, 2019.

\bibitem[Betancourt(2017)]{betancourt2017}
Michael Betancourt.
\newblock {A Conceptual Introduction to Hamiltonian Monte Carlo}.
\newblock \emph{arXiv preprint arXiv:1701.02434}, 2017.
\newblock URL \url{https://arxiv.org/abs/1701.02434}.

\bibitem[Neal(2011)]{neal2011}
Radford~M. Neal.
\newblock {MCMC Using Hamiltonian dynamics}.
\newblock \emph{Handbook of Markov Chain Monte Carlo}, 2:\penalty0 113--162, 2011.
\newblock \doi{10.1201/b10905}.

\bibitem[Bales et~al.(2019)Bales, Pourzanjani, Vehtari, and Petzold]{bales2019}
Ben Bales, Arya Pourzanjani, Aki Vehtari, and Linda Petzold.
\newblock {Selecting the Metric in Hamiltonian Monte Carlo}, 2019.

\bibitem[Kucukelbir et~al.(2017)Kucukelbir, Tran, Ranganath, Gelman, and Blei]{kucukelbir2017}
Alp Kucukelbir, Dustin Tran, Rajesh Ranganath, Andrew Gelman, and David~M. Blei.
\newblock {Automatic Differentiation Variational Inference}.
\newblock \emph{Journal of Machine Learning Research}, 18\penalty0 (14):\penalty0 1--45, 2017.

\bibitem[Zhang et~al.(2022)Zhang, Carpenter, Gelman, and Vehtari]{zhang2022}
Lu~Zhang, Bob Carpenter, Andrew Gelman, and Aki Vehtari.
\newblock {Pathfinder: Parallel quasi-Newton variational inference}.
\newblock \emph{Journal of Machine Learning Research}, 23\penalty0 (306):\penalty0 1--49, 2022.

\bibitem[Margossian et~al.(2020)Margossian, Vehtari, Simpson, and Agrawal]{margossian2020}
Charles Margossian, Aki Vehtari, Daniel Simpson, and Raj Agrawal.
\newblock {Hamiltonian {M}onte {C}arlo using an adjoint-differentiated {L}aplace approximation: {B}ayesian inference for latent {G}aussian models and beyond}.
\newblock In H.~Larochelle, M.~Ranzato, R.~Hadsell, M.F. Balcan, and H.~Lin, editors, \emph{Advances in {N}eural {I}nformation {P}rocessing {S}ystems}, volume~33, pages 9086--9097. Curran Associates, Inc., 2020.

\bibitem[Carpenter et~al.(2017)Carpenter, Gelman, Hoffman, Lee, Goodrich, Betancourt, Brubaker, Guo, Li, and Riddell]{carpenter2017}
Bob Carpenter, Andrew Gelman, Matthew~D. Hoffman, Daniel Lee, Ben Goodrich, Michael Betancourt, Marcus Brubaker, Jiqiang Guo, Peter Li, and Allen Riddell.
\newblock {Stan: A Probabilistic Programming Language}.
\newblock \emph{Journal of Statistical Software}, 76\penalty0 (1):\penalty0 1--29, 2017.
\newblock ISSN 1548-7660.
\newblock \doi{10.18637/jss.v076.i01}.

\bibitem[{Stan Development Team}(2025)]{stan2025}
{Stan Development Team}.
\newblock {Stan Reference Manual. v2.37.0.}, 2025.
\newblock URL \url{http://mc-stan.org}.

\bibitem[Abril-Pla et~al.(2023)Abril-Pla, Andreani, Carroll, Dong, Fonnesbeck, Kochurov, Kumar, Lao, Luhmann, Martin, Osthege, Vieira, Wiecki, and Zinkov]{abril2023}
Oriol Abril-Pla, Virgile Andreani, Colin Carroll, Larry Dong, Christopher~J. Fonnesbeck, Maxim Kochurov, Ravin Kumar, Junpeng Lao, Christian~C. Luhmann, Osvaldo~A. Martin, Michael Osthege, Ricardo Vieira, Thomas Wiecki, and Robert Zinkov.
\newblock {PyMC: a modern, and comprehensive probabilistic programming framework in Python}.
\newblock \emph{PeerJ Computer Science}, 9:\penalty0 e1516, 2023.

\bibitem[Kristensen et~al.(2016)Kristensen, Nielsen, Berg, Skaug, and Bell]{kristensen2016}
Kasper Kristensen, Anders Nielsen, Casper~W. Berg, Hans Skaug, and Bradley~M. Bell.
\newblock {TMB: Automatic differentiation and Laplace approximation}.
\newblock \emph{Journal of Statistical Software}, 70\penalty0 (5):\penalty0 21, 2016.
\newblock \doi{10.18637/jss.v070.i05}.

\bibitem[Brooks et~al.(2017)Brooks, Kristensen, van Benthem, Magnusson, Berg, Nielsen, Skaug, Maechler, and Bolker]{brooks2017}
Mollie~E. Brooks, Kasper Kristensen, Koen~J. van Benthem, Arni Magnusson, Casper~W. Berg, Anders Nielsen, Hans~J. Skaug, Martin Maechler, and Benjamin~M. Bolker.
\newblock {glmmTMB balances speed and flexibility among packages for zero-inflated generalized linear mixed modeling}.
\newblock \emph{R Journal}, 9\penalty0 (2):\penalty0 378--400, 2017.
\newblock \doi{10.32614/Rj-2017-066}.

\bibitem[Thorson et~al.(2025{\natexlab{a}})Thorson, Anderson, Goddard, and Rooper]{thorson2025b}
James~T. Thorson, Sean~C. Anderson, Pamela Goddard, and Christopher~N. Rooper.
\newblock {tinyVAST: R Package With an Expressive Interface to Specify Lagged and Simultaneous Effects in Multivariate Spatio-Temporal Models}.
\newblock \emph{Global Ecology and Biogeography}, 34\penalty0 (4):\penalty0 e70035, 2025{\natexlab{a}}.
\newblock \doi{https://doi.org/10.1111/geb.70035}.

\bibitem[Anderson et~al.(2025)Anderson, Ward, English, Barnett, and Thorson]{anderson2025}
Sean~C. Anderson, Eric~J. Ward, Philina~A. English, Lewis A.~K. Barnett, and James~T. Thorson.
\newblock {sdmTMB}: An {R} package for fast, flexible, and user-friendly generalized linear mixed effects models with spatial and spatiotemporal random fields.
\newblock \emph{Journal of Statistical Software}, 115\penalty0 (2):\penalty0 1–46, 2025.
\newblock \doi{10.18637/jss.v115.i02}.
\newblock URL \url{https://www.jstatsoft.org/index.php/jss/article/view/v115i02}.

\bibitem[Kellner et~al.(2023)Kellner, Smith, Royle, Kery, Belant, and Chandler]{kellner2023}
Kenneth~F. Kellner, Adam~D. Smith, J.~Andrew Royle, Marc Kery, Jerrold~L. Belant, and Richard~B. Chandler.
\newblock {The unmarked R package: Twelve years of advances in occurrence and abundance modelling in ecology}.
\newblock \emph{Methods in Ecology and Evolution}, 14\penalty0 (6):\penalty0 1408--1415, 2023.
\newblock \doi{https://doi.org/10.1111/2041-210X.14123}.

\bibitem[Rue and Held(2005)]{rue2005}
Havard Rue and Leonhard Held.
\newblock \emph{{Gaussian Markov random fields: theory and applications}}.
\newblock Chapman and Hall/CRC, 2005.

\bibitem[Lindgren et~al.(2011)Lindgren, Rue, and Lindstrom]{lindgren2011}
Finn Lindgren, Havard Rue, and Johan Lindstrom.
\newblock {An explicit link between Gaussian fields and Gaussian Markov random fields: the stochastic partial differential equation approach}.
\newblock \emph{Journal of the Royal Statistical Society Series B-Statistical Methodology}, 73\penalty0 (4):\penalty0 423--498, 2011.
\newblock ISSN 1369-7412.
\newblock \doi{10.1111/j.1467-9868.2011.00777.x}.

\bibitem[Kristensen(2025)]{rtmb2025}
Kasper Kristensen.
\newblock \emph{{RTMB: 'R' Bindings for 'TMB'}}, 2025.
\newblock URL \url{https://github.com/kaskr/RTMB}.
\newblock R package version 1.7.

\bibitem[Betancourt(2016)]{betancourt2016}
Michael Betancourt.
\newblock {Identifying the optimal integration time in Hamiltonian Monte Carlo}.
\newblock \emph{arXiv preprint arXiv:1601.00225}, 2016.

\bibitem[Bates et~al.(2025)Bates, Maechler, and Jagan]{matrix2025}
Douglas Bates, Martin Maechler, and Mikael Jagan.
\newblock \emph{{Matrix: Sparse and Dense Matrix Classes and Methods}}, 2025.
\newblock URL \url{https://CRAN.R-project.org/package=Matrix}.
\newblock R package version 1.7-3.

\bibitem[Skaug and Fournier(2006)]{skaug2006}
Hans~J. Skaug and David~A. Fournier.
\newblock {Automatic approximation of the marginal likelihood in non-Gaussian hierarchical models}.
\newblock \emph{Computational Statistics and Data Analysis}, 51\penalty0 (2):\penalty0 699--709, 2006.
\newblock \doi{10.1016/j.csda.2006.03.005}.

\bibitem[Kass and Steffey(1989)]{kass1989}
Robert~E. Kass and David Steffey.
\newblock {Approximate Bayesian-inference in conditionally independent hierarchical-models (parametric empirical Bayes models)}.
\newblock \emph{Journal of the American Statistical Association}, 84\penalty0 (407):\penalty0 717--726, 1989.
\newblock \doi{10.2307/2289653}.

\bibitem[Bell and Kristensen(2021)]{bell2021}
Bradley~M. Bell and Kasper Kristensen.
\newblock {Computing sparse Jacobians and Hessians using algorithmic differentiation}.
\newblock \emph{arXiv preprint arXiv:2111.05207}, 2021.
\newblock URL \url{https://arxiv.org/abs/2111.05207}.

\bibitem[Zheng and Cadigan(2021)]{zheng2021}
Nan Zheng and Noel Cadigan.
\newblock {Frequentist delta-variance approximations with mixed-effects models and TMB}.
\newblock \emph{Computational Statistics and Data Analysis}, 160:\penalty0 107227, 2021.
\newblock \doi{https://doi.org/10.1016/j.csda.2021.107227}.

\bibitem[Johnson(2025)]{johnson2025}
Andrew~R. Johnson.
\newblock \emph{{StanEstimators: Estimate parameters for arbitrary R functions using 'Stan'}}, 2025.
\newblock URL \url{https://github.com/andrjohns/StanEstimators}.
\newblock R package version 0.1.2.9000.

\bibitem[McCann(1995)]{mccann1995}
Robert~J. McCann.
\newblock {Existence and uniqueness of monotone measure-preserving maps}.
\newblock \emph{Duke Mathematical Journal}, 80\penalty0 (2):\penalty0 309--323, 1995.
\newblock \doi{10.1215/S0012-7094-95-08013-2}.

\bibitem[Craig(2016)]{craig2016}
Katy Craig.
\newblock {The exponential formula for the Wasserstein metric}.
\newblock \emph{Esaim-Control Optimisation and Calculus of Variations}, 22\penalty0 (1):\penalty0 169--187, 2016.
\newblock \doi{10.1051/cocv/2014069}.

\bibitem[Bürkner et~al.(2025)Bürkner, Gabry, Kay, and Vehtari]{buerkner2025}
Paul-Christian Bürkner, Jonah Gabry, Matthew Kay, and Aki Vehtari.
\newblock {posterior: Tools for Working with Posterior Distributions}, 2025.
\newblock URL \url{https://mc-stan.org/posterior/}.
\newblock R package version 1.6.1.

\bibitem[Vehtari et~al.(2021)Vehtari, Gelman, Simpson, Carpenter, and Bürkner]{vehtari2021}
Aki Vehtari, Andrew Gelman, Daniel Simpson, Bob Carpenter, and Paul-Christian Bürkner.
\newblock {Rank-normalization, folding, and localization: An improved $\hat{R}$ for assessing convergence of MCMC}.
\newblock \emph{Bayesian Analysis}, 16\penalty0 (2):\penalty0 667--718, 2021.
\newblock \doi{10.1214/20-ba1221}.

\bibitem[Monnahan and Kristensen(2018)]{monnahan2018}
Cole~C. Monnahan and Kasper Kristensen.
\newblock {No-U-turn sampling for fast {B}ayesian inference in {ADMB} and {TMB}: {I}ntroducing the adnuts and tmbstan {R} packages}.
\newblock \emph{Plos One}, 13\penalty0 (5):\penalty0 e0197954, 2018.
\newblock \doi{10.1371/journal.pone.0197954}.

\bibitem[Magnusson et~al.(2024)Magnusson, Torgander, Bürkner, Zhang, Carpenter, and Vehtari]{magnusson2024}
Mans Magnusson, Jakob Torgander, Paul-Christian Bürkner, Lu~Zhang, Bob Carpenter, and Aki Vehtari.
\newblock {posteriordb: Testing, benchmarking and developing Bayesian inference algorithms}.
\newblock \emph{arXiv preprint arXiv:2407.04967}, 2024.

\bibitem[Lindgren and Rue(2015)]{lindgren2015}
Finn Lindgren and Havard Rue.
\newblock {Bayesian spatial modelling with {R-INLA}}.
\newblock \emph{{Journal of Statistical Software}}, 63\penalty0 (19):\penalty0 1--25, 2015.

\bibitem[Gelman et~al.(2020)Gelman, Vehtari, Simpson, Margossian, Carpenter, Yao, Kennedy, Gabry, Bürkner, and Modrák]{gelman2020}
Andrew Gelman, Aki Vehtari, Daniel Simpson, Charles~C. Margossian, Bob Carpenter, Yuling Yao, Lauren Kennedy, Jonah Gabry, Paul-Christian Bürkner, and Martin Modrák.
\newblock {Bayesian workflow}.
\newblock \emph{arXiv preprint arXiv:2011.01808}, 2020.

\bibitem[Papaspiliopoulos et~al.(2007)Papaspiliopoulos, Roberts, and Skold]{papaspiliopoulos2007}
Omiros Papaspiliopoulos, Gareth~O. Roberts, and Martin Skold.
\newblock {A general framework for the parametrization of hierarchical models}.
\newblock \emph{Statistical Science}, 22\penalty0 (1):\penalty0 59--73, 2007.
\newblock \doi{10.1214/088342307000000014}.

\bibitem[Osgood-Zimmerman and Wakefield(2023)]{osgood2023}
Aaron Osgood-Zimmerman and Jon Wakefield.
\newblock {A statistical review of Template Model Builder: A flexible tool for spatial modelling}.
\newblock \emph{{International Statistical Review}}, 91\penalty0 (2):\penalty0 318--342, 2023.

\bibitem[Gabry and Mahr(2025)]{gabry2025}
Jonah Gabry and Tristan Mahr.
\newblock {bayesplot: Plotting for Bayesian models}, 2025.
\newblock URL \url{https://mc-stan.org/bayesplot/}.
\newblock R package version 1.14.0.

\bibitem[Schuhmacher et~al.(2024)Schuhmacher, Bähre, Bonneel, Gottschlich, Hartmann, Heinemann, Schmitzer, and Schrieber]{schumacher2020}
Dominic Schuhmacher, Björn Bähre, Nicolas Bonneel, Carsten Gottschlich, Valentin Hartmann, Florian Heinemann, Bernhard Schmitzer, and Jörn Schrieber.
\newblock \emph{{{transport}: computation of optimal transport plans and Wasserstein distances}}, 2024.
\newblock R package version 0.15-4.

\bibitem[Jonsen et~al.(2023)Jonsen, Grecian, Phillips, Carroll, McMahon, Harcourt, Hindell, and Patterson]{jonsen2023animotum}
Ian~D. Jonsen, W.~James Grecian, Lachlan Phillips, Gemma Carroll, Clive McMahon, Robert~G. Harcourt, Mark~A. Hindell, and Toby~A. Patterson.
\newblock {aniMotum, an R package for animal movement data: Rapid quality control, behavioural estimation and simulation}.
\newblock \emph{Methods in Ecology and Evolution}, 14\penalty0 (3):\penalty0 806--816, 2023.

\bibitem[Vetter et~al.(2025)Vetter, Møller, Thygesen, Bacher, and Madsen]{vetter2025}
Phillip~Brinck Vetter, Jan~K. Møller, Uffe Thygesen, Peder Bacher, and Henrik Madsen.
\newblock \emph{{ctsmTMB: Continuous Time Stochastic Modelling using Template Model Builder}}, 2025.
\newblock R Package Version 1.0.

\bibitem[Thorson et~al.(2024)Thorson, III, Essington, and Large]{thorson2024}
James~T. Thorson, Albert G.~Andrews III, Timothy~E. Essington, and Scott~I. Large.
\newblock {Dynamic structural equation models synthesize ecosystem dynamics constrained by ecological mechanisms}.
\newblock \emph{Methods in Ecology and Evolution}, 15\penalty0 (4):\penalty0 744--755, 2024.
\newblock \doi{10.1111/2041-210x.14289}.

\bibitem[Thorson et~al.(2025{\natexlab{b}})Thorson, Kristensen, Aydin, Gaichas, Kimmel, McHuron, Nielsen, Townsend, and Whitehouse]{thorson2025}
James~T. Thorson, Kasper Kristensen, Kerim~Y. Aydin, Sarah~K. Gaichas, David~G. Kimmel, Elizabeth~A. McHuron, Jens~M. Nielsen, Howard Townsend, and George~A. Whitehouse.
\newblock {The Benefits of hierarchical ecosystem models: Demonstration using EcoState, a new state-space mass-balance model}.
\newblock \emph{Fish and Fisheries}, 26\penalty0 (2):\penalty0 203--218, 2025{\natexlab{b}}.

\bibitem[Magnusson and Maunder(2025)]{magnusson2025}
Arni Magnusson and Mark Maunder.
\newblock \emph{{fishgrowth: Fit Growth Curves to Fish Data}}, 2025.
\newblock R package version 1.0.4.

\bibitem[Niku et~al.(2019)Niku, Hui, Taskinen, and Warton]{niku2019}
Jenni Niku, Francis K.~C. Hui, Sara Taskinen, and David~I. Warton.
\newblock {gllvm: Fast analysis of multivariate abundance data with generalized linear latent variable models in R}.
\newblock \emph{Methods in Ecology and Evolution}, 10\penalty0 (12):\penalty0 2173--2182, 2019.
\newblock \doi{https://doi.org/10.1111/2041-210X.13303}.

\bibitem[Michelot(2025)]{michelot2025}
Theo Michelot.
\newblock {hmmTMB: Hidden Markov models with flexible covariate effects in R}.
\newblock \emph{Journal of Statistical Software}, 114\penalty0 (5):\penalty0 1--45, 2025.
\newblock \doi{10.18637/jss.v114.i05}.

\bibitem[Mews et~al.(2025)Mews, Koslik, and Langrock]{mews2025}
Sina Mews, Jan-Ole Koslik, and Roland Langrock.
\newblock {How to build your latent Markov model: The role of time and space}.
\newblock \emph{Statistical Modelling}, 25\penalty0 (6):\penalty0 481--507, 2025.
\newblock \doi{10.1177/1471082X251355681}.

\bibitem[Laake et~al.(2013)Laake, Johnson, and Conn]{laake2013}
Jeff~L. Laake, Devin~S. Johnson, and Paul~B. Conn.
\newblock {marked: an R package for maximum likelihood and Markov Chain Monte Carlo analysis of capture–recapture data}.
\newblock \emph{Methods in Ecology and Evolution}, 4\penalty0 (9):\penalty0 885--890, 2013.
\newblock \doi{https://doi.org/10.1111/2041-210X.12065}.

\bibitem[Bove et~al.(2025)Bove, Stock, Li, He, and Adebola]{bove2025}
Daniel~Sabanes Bove, Christian Stock, Liming Li, Jun He, and Kayode Adebola.
\newblock \emph{{{mmrm}: Mixed Models for Repeated Measures}}, 2025.
\newblock R package version 0.3.16.

\bibitem[Thorson and van~der Bijl(2023)]{thorson2023phylosem}
James~T. Thorson and Wouter van~der Bijl.
\newblock {phylosem: A fast and simple R package for phylogenetic inference and trait imputation using phylogenetic structural equation models}.
\newblock \emph{Journal of Evolutionary Biology}, 36\penalty0 (10):\penalty0 1357--1364, 2023.

\bibitem[Wahl(2025)]{wahl2025}
Jens Wahl.
\newblock \emph{{stochvolTMB: Likelihood Estimation of Stochastic Volatility Models}}, 2025.
\newblock R package version 0.3.0.

\bibitem[Nielsen and Berg(2014)]{nielsen2014}
Anders Nielsen and Casper~W. Berg.
\newblock {Estimation of time-varying selectivity in stock assessments using state-space models}.
\newblock \emph{Fisheries Research}, 158:\penalty0 96--101, 2014.
\newblock \doi{10.1016/j.fishres.2014.01.014}.

\bibitem[Galanos(2025)]{galanos2025}
Alexios Galanos.
\newblock \emph{{tsissm: Linear innovations state space unobserved components model}}, 2025.
\newblock R package version 1.0.2.

\bibitem[Thorson(2019)]{thorson2019}
James~T. Thorson.
\newblock {Guidance for decisions using the Vector Autoregressive Spatio-Temporal (VAST) package in stock, ecosystem, habitat and climate assessments}.
\newblock \emph{Fisheries Research}, 210:\penalty0 143--161, 2019.
\newblock \doi{https://doi.org/10.1016/j.fishres.2018.10.013}.

\bibitem[Stock and Miller(2021)]{stock2021}
Brian~C. Stock and Timothy~J. Miller.
\newblock {The Woods Hole Assessment Model (WHAM): A general state-space assessment framework that incorporates time- and age-varying processes via random effects and links to environmental covariates}.
\newblock \emph{Fisheries Research}, 240:\penalty0 105967, 2021.
\newblock \doi{https://doi.org/10.1016/j.fishres.2021.105967}.

\bibitem[Descamps et~al.(2016)Descamps, Tarroux, Cherel, Delord, Godo, Kato, Krafft, Lorentsen, Ropert-Coudert, Skaret, and Varpe]{descamps2016}
Sebastien Descamps, Arnaud Tarroux, Yves Cherel, Karine Delord, Olav~R. Godo, Akiko Kato, Bjorn~A. Krafft, Svein-Hakon Lorentsen, Yan Ropert-Coudert, Georg Skaret, and Oystein Varpe.
\newblock {Data from: At-sea distribution and prey selection of Antarctic petrels and commercial krill fisheries}, 2016.
\newblock URL \url{http://dx.doi.org/10.5441/001/1.q4gn4q56}.

\bibitem[Monnahan et~al.(2024)Monnahan, Ferriss, Shotwell, Oyafuso, Levine, Thorson, Rogers, Sullivan, and Champagnat]{monnahan2024}
Cole~C. Monnahan, Bridget~E. Ferriss, S.~Kalei Shotwell, Zack Oyafuso, Mike Levine, James~T. Thorson, Lauren Rogers, Jane Sullivan, and Juliette Champagnat.
\newblock {Assessment of the walleye pollock stock in the Gulf of Alaska. In Stock Assessment and Fishery Evaluation Report for Groundfish Resources of the Gulf of Alaska.}
\newblock Technical report, Prepared by the Gulf of Alaska Groundfish Plan Team, North Pacific Fishery Management Council, Anchorage, AK., 2024.

\bibitem[Korner-Nievergelt et~al.(2015)Korner-Nievergelt, Roth, von Felten, Guelat, Almasi, and Korner-Nievergelt]{korner2015}
Franzi Korner-Nievergelt, Tobias Roth, Stefanie von Felten, Jerome Guelat, Bettina Almasi, and Pius Korner-Nievergelt.
\newblock \emph{{Bayesian data analysis in ecology using linear models with {R}, {BUGS}, and {Stan}: including comparisons to frequentist statistics}}.
\newblock Academic Press, 2015.
\newblock ISBN 0128016787.

\bibitem[Bolker et~al.(2013)Bolker, Gardner, Maunder, Berg, Brooks, Comita, Crone, Cubaynes, Davies, de~Valpine, Ford, Gimenez, Kery, Kim, Lennert-Cody, Magnusson, Martell, Nash, Nielsen, Regetz, Skaug, and Zipkin]{bolker2013}
Benjamin~M. Bolker, Beth Gardner, Mark Maunder, Casper~W. Berg, Mollie Brooks, Lizzie Comita, Elizabeth Crone, Sarah Cubaynes, Tyrell Davies, Perry de~Valpine, John Ford, Olivier Gimenez, Marc Kery, Eun~Jung Kim, Cleridy Lennert-Cody, Arni Magnusson, Steven Martell, John Nash, Anders Nielsen, Jim Regetz, Hans Skaug, and Elise Zipkin.
\newblock {Strategies for fitting nonlinear ecological models in R, AD Model Builder, and BUGS}.
\newblock \emph{Methods in Ecology and Evolution}, 4\penalty0 (6):\penalty0 501--512, 2013.
\newblock ISSN 2041-210X.
\newblock \doi{10.1111/2041-210X.12044}.

\end{thebibliography}

\appendix
% Add 'S' prefix to figure and table numbering
\renewcommand{\thefigure}{S\arabic{figure}}
\renewcommand{\thetable}{S\arabic{table}}

\setcounter{figure}{0}
\setcounter{table}{0}

\section{Supplementary sections}

\subsection{Sensitivity to initialization for approximate methods}\label{sec:pfinits}
In the main text we initialized Pathfinder and \package{TMB} from samples from the posterior. We arrived at this by first we testing the sensitivity to the choice of initialization on the simple \model{schools} model. We randomly initialized all parameters at $U(-2,2)$, $U(-1,1)$, or from the reference posterior, as well as two static vectors: the conditional mode $\hat{q}$ and all parameters at 0. We ran 20 replicates for each and found similar run time performance across the initialization methods, with precision sampling being about 100 times faster than Pathfinder (\figref{fig:pfinits}). Interestingly, Pathfinder’s approximate samples had more variation in the Wasserstein distance, particularly for draws initialized from the posterior, and the lowest variance when initialized at 0. Results for precision sampling were very consistent because marginal optimization was able to find the same mode and so all variation in the Wasserstein distance was from the random sample generation. This suggests that \package{TMB} is less sensitive to initialization than Pathfinder. We used posterior samples as this seemed a straightforward compromise, however in practice posterior samples are not available and so an alterative approach would be used.

\subsection{Automatic selection of a preconditioner type}\label{sec:auto}
The best preconditioner (i.e., diag, dense, or sparse) for a given model will depend on many factors. Many models will have no marginal mode and hence no $Q$, $Q$ may be available but have no sparsity or minimal global correlations (small condition number), or have no random effects and thus only $\Sigma$ is available. These situations imply different approaches to preconditioning, but fortunately this information is typically available before starting NUTS sampling and so we propose a heuristic method to automatically select an approach.

We use the following scheme to automatically select the decorrelation type:
\begin{itemize}
\item	If neither $Q$ nor $\Sigma$ is available, use the Stan default of estimating a diagonal mass matrix.
\item	If correlations are low, use a diagonal matrix to descale the posterior only, because the dense and sparse preconditioners would only add computational overhead.
\item	If correlations are high (a maximum of pairwise correlations higher than 0.3), use a dense or sparse matrix, depending on which gradient is faster to compute (i.e., equation 2 or 3; Fig. 3)).
\end{itemize}

Since the preconditioning is applied externally to Stan, the mass matrix $M$ used internally by Stan is applied to the preconditioned posterior. In theory, if $Q^{-1}$ approximates the posterior geometry sufficiently well then no further decorrelation or descaling would be useful, so Stan's mass matrix adaptation can be disabled, and consequently a much shorter warmup period of 150 warmup samples used (see \secref{sec:warmup} in Appendix A). This can be diagnosed post-hoc by comparing estimates of marginal variances and pairwise parameter correlations of the posterior to those implied by $Q$, with large mismatches indicating issues with the approximation and likely degradation of SNUTS efficiency.  We show an example using the \model{wildf} model in the main text of when this can further improve sampling efficiency

Importantly, our automatic selection procedure adds additional computational cost to a NUTS analysis, specifically marginal optimization, estimation of $Q$, inversion of $Q^{-1}=\Sigma$ to get correlations, and if correlations are high the Cholesky factorization $\text{chol}(\Sigma)$ and $\text{chol}(Q_P)$ to compare dense and sparse gradient costs. We hypothesize that this computational overhead is minimal compared to the time for a full NUTS analysis, and also test this claim below.

\subsection{Determining a sufficient warmup period}\label{sec:warmup}
The default and common configuration of NUTS in Stan is to adapt both the step size and a diagonal mass matrix during the warmup phase, using a series of expanding windows \citep{stan2025}. This procedure is designed to be robust across various model types, but requires a substantial warmup period, typically 1000 warmup iterations or 50\% of the total iterations. Here we claim that $Q$ represents both the marginal variances and global correlations and when well-approximated, adaptation of the mass matrix is unnecessary and consequently the warmup period need only be long enough to move to the typical set and adapt the step size.

We tested this claim by running typical adaptation in Stan for SNUTS chains and examining the behavior of the mass matrix adaptation on the step size across the expanding windows to see how long warmup needs to stabilize estimates of step size. We initialized each of 10 chains from a random draw from $Q$ as described in the main text (precision sampling).  We found stable step sizes for all models even after the first window (\figref{fig:warmup}), suggesting that diagonal mass matrix adaptation is not meaningfully rescaling the posterior, except perhaps for the \model{wildf} model (see Results section in the main text). We therefore disabled Stan's mass matrix adaptation and used 1250 total iterations with 150 warmup iterations for all SNUTS analyses, in contrast to NUTS analyses which use Stan defaults of 2000 and 1000, respectively.

\subsection{Example usage of the \package{SparseNUTS} R package}

The main function is \func{sample\_snuts(obj)} which takes \package{TMB}'s dynamic list object (\texttt{obj}) and other optional arguments, and returns a member of a custom R S3 class \texttt{tmbfit} which is a list containing important outputs, such as the conditional mode $\hat{q}$, $Q$, and parameter correlations, as well as the posterior samples and metadata about the SNUTS analysis. The user can specify standard NUTS inputs such as number of chains, number of warmup and samples per chain, a control list to specify e.g., the maximum tree depth and target acceptance rates, but also SNUTS specific options like the type of preconditioner via the \texttt{metric} argument (default of ``auto''), and controls to override default behavior for marginal optimization and calculation of $Q$ internally.

We demonstrate basic usage on a glmmTMB negative binomial regression with random sites as follows (see also \textit{salamanders} in Tables 1, \ref{tab:mods2}).

\begin{Verbatim}
library(glmmTMB)
data(Salamanders)
mod <- glmmTMB(count~spp * mined + (1|site), Salamanders, family="nbinom2")

library(SparseNUTS)
fit <- sample_snuts(obj=mod$obj)
print(fit)
\end{Verbatim}
This results in the following console output.
\begin{Verbatim}
Model 'glmmTMB' has 39 pars, and was fit using NUTS with a 'dense' metric
4 chain(s) of 1150 total iterations (150 warmup) were used
Average run time per chain was 5.5 seconds
Minimum ESS=1284.6 (32.11%), and maximum Rhat=1.003
There were 0 divergences after warmup
\end{Verbatim}
In this example, $Q$ is 44.5\% sparse and the highest pairwise correlation was 0.96 and we can see that a dense preconditioner and a short warmup period without mass matrix adaptation was automatically chosen.  \SparseNUTS has a set of member functions to allow the user perform standard tasks in a Bayesian analysis. One unique aspect of SNUTS is that has estimates from an approximate (i.e., $Q$) and full MCMC sampling. These can be compared to evaluate the accuracy of the approximation to, for instance, gauge whether an approximation is sufficient during early model development or to better understand why the preconditioning may not improve sampling efficiency. Specifically, \SparseNUTS has a \texttt{pairs} member function to show pairwise plots of the posterior overlaid with the 95\% confidence ellipses from $Q^{-1}$ centered at $\hat{q}$ , a function \texttt{plot\_uncertainties} to compare asymptotic and posterior estimates of marginal variances, and \texttt{plot\_marginals} to plot histograms.

Posterior samples can be extracted into an R data frame via \texttt{as.data.frame()} and put back into the \package{TMB} object's \texttt{report} function to get posterior samples from derived quantities. This output can also be plugged into external Bayesian R packages for other components of a typical Bayesian workflow like model selection via PSIS-LOO \citep{vehtari2021}, specialized MCMC plotting via \texttt{bayesplot} \citep{gabry2025}, and calculating and visualization of posterior predictive checks. The package vignettes provide worked examples of these and other capabilities of the package: \url[{https://cole-monnahan-noaa.github.io/SparseNUTS/}.

\subsection{Details of testing Pathfinder vs TMB}
We tested the computation time for both approximation approaches. For Pathfinder run time includes optimization, generating random normal samples, and evaluating evidence of lower bounds. We used four paths (the default), but did not run them in parallel for simplicity of comparison. The computational cost for  precision sampling is optimization of the marginal posterior, calculation of $Q$ (equation 4), inversion of $Q$, and random normal sample generation. For large, highly sparse $Q$ matrices sample generation likely would be more efficient if done directly from $Q$ (i.e,. without inverting $Q$; algorithm 2.4 of \cite{rue2005}) but that was not explored here as this computational aspect was minimal in our case studies. Preliminary evaluations on the set of case studies (Table~1) led to many numerical issues and unstable estimates for Pathfinder when initialized from dispersed points, so we used posterior draws for all models.

To calculate the 1-Wasserstein distance we used the \func{wasserstein} function from the \package{transport} R package \citep{schumacher2020}. An estimated Wasserstein distance represents how much one empirical distribution would need to be distorted to match another one, with larger values indicating larger differences, and values closer to zero more similar distributions. We computed the  distance of 1000 approximate samples from both algorithms to 10,000 NUTS samples, assumed to represent the true posterior distribution.
\clearpage

\section{Supplementary figures and tables}

\begin{figure}[ht]
    \centering
    \includegraphics[scale=.6]{"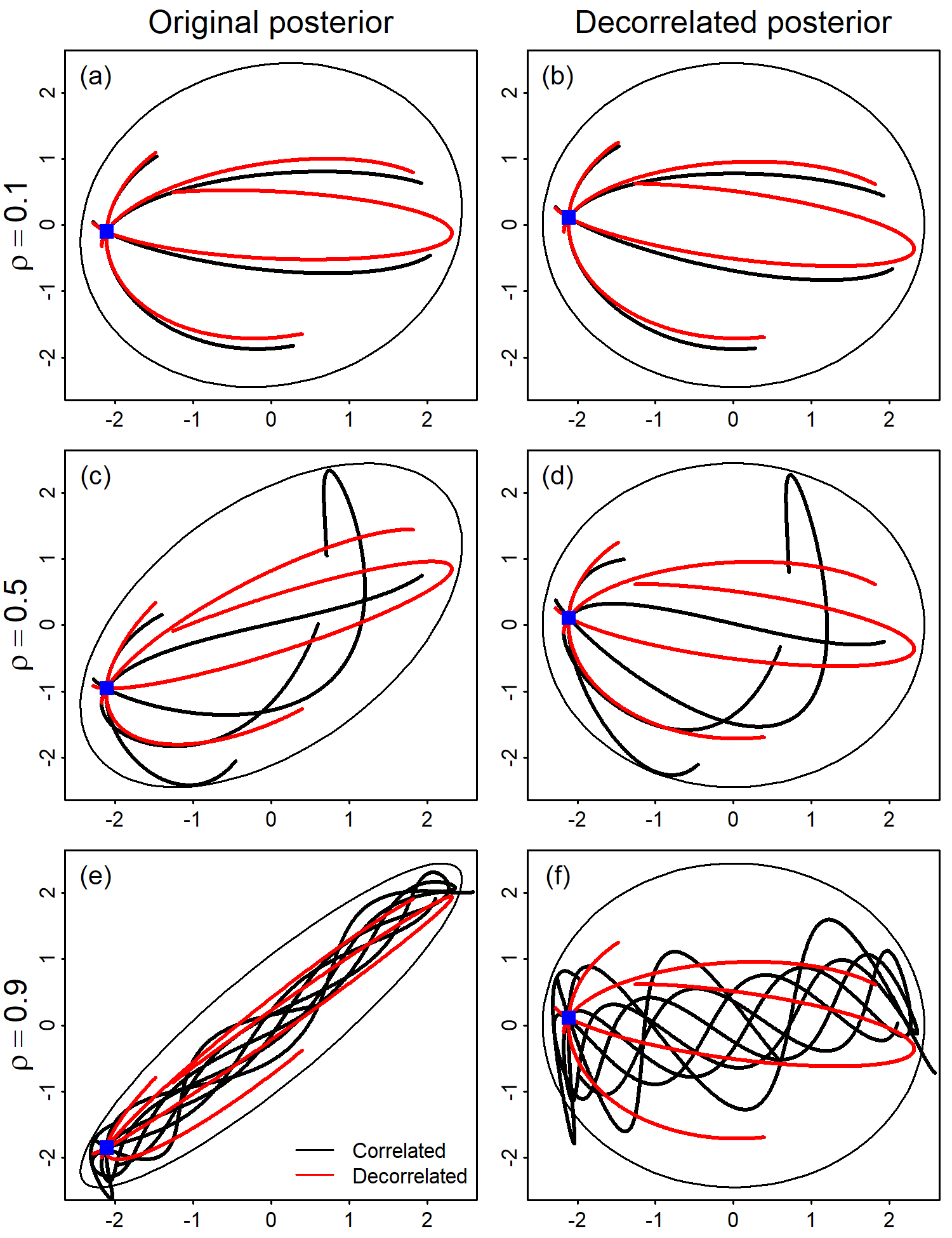"}
    \caption{{\bfseries Example NUTS trajectories.}  The plots show a bivariate normal posterior density with unit variances but different levels of correlation $\rho$ (rows). Shown are 5 trajectories (lines) initiated from the same point (blue square) but with different random momenta. The trajectories are calculated with and without a dense preconditioner (colors), which are shown either on the original posterior ($q$; left) or preconditioned posterior ($q'$; right). The preconditioned trajectories are identical between panels}
    \label{fig:trajectories}
\end{figure}

\begin{figure}[ht]
\includegraphics[scale=.7,center]{"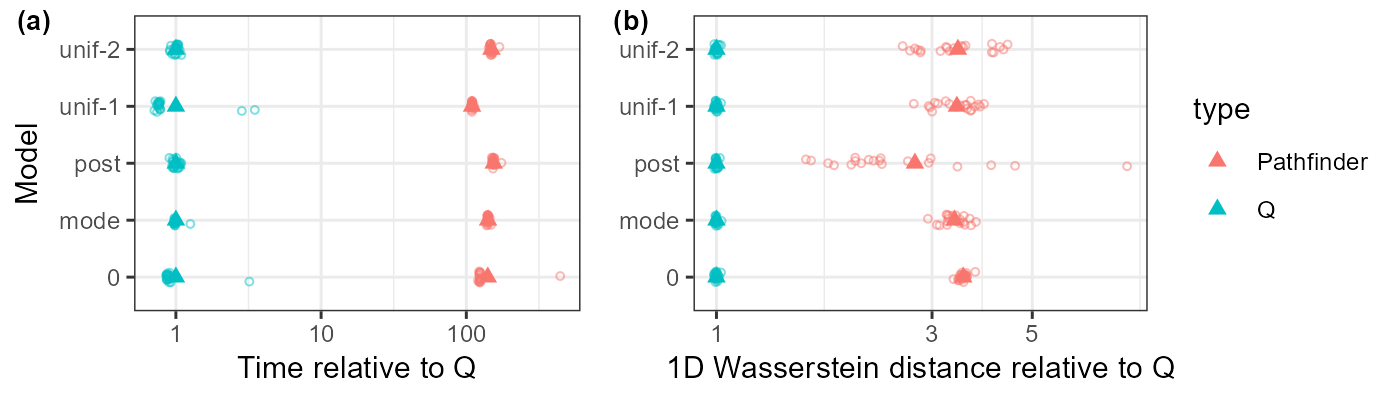"}
\caption{{\bfseries Further Pathfinder evaluations.} The plots compare the approximation performance of the  precision ($Q$) approximation to the Pathfinder algorithm for the \model{schools} model across five different methods for generating initial values. Methods include random draws from U(-2,2) (‘unif-2’) or U(-1,1) (‘unif-1’)), random draws from the posterior (‘post’), the empirical Bayesian conditional mode from \package{TMB} (‘mode’) and all parameters at zero (‘0’). Each of the 20 point represents a single analysis with a different random initial parameter vector (used for both methods, if appropriate) and random seed for sample generation. Means are given by filled triangles. Pathfinder used 4 serial paths, while $Q$ optimized once and generated samples.
}
\label{fig:pfinits}
\end{figure}

\begin{figure}[ht]
\includegraphics[scale=.7]{"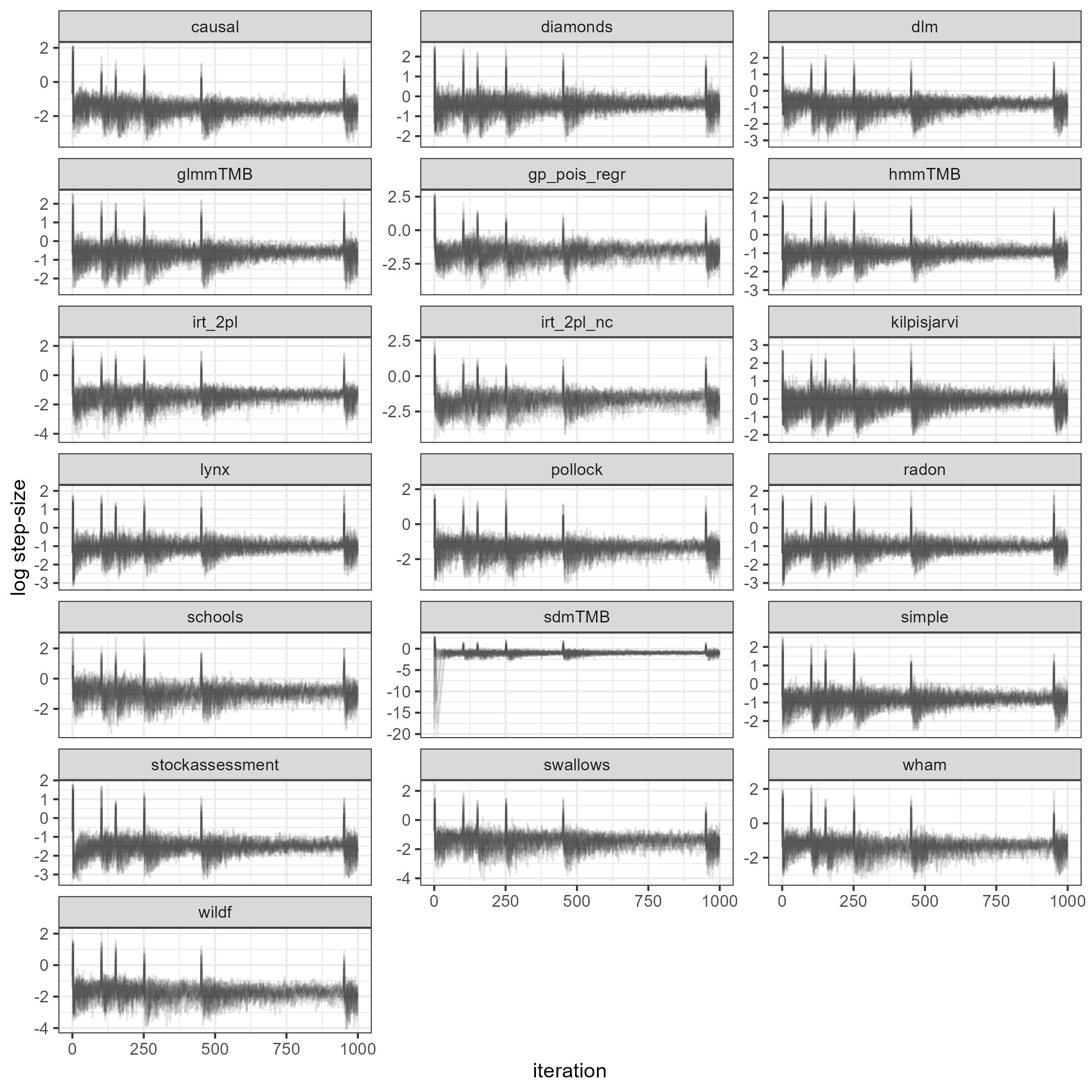"}
\caption{{\bfseries Adaptation results}. The plots show the log of the NUTS step size for 1000 warmup iterations across 10 chains, initialized with random values using precision sampling. Stan’s adaptation scheme updates the mass matrix (\texttt{diag\_e} in this case) at expanding intervals after which the step size will update accordingly. Each model uses the ``auto'' setting to automatically determine the preconditioner, so the mass matrix is not expected to change significantly across the warmup iterations when the posterior is well approximated by $Q$.}
\label{fig:warmup}
\end{figure}

\begin{figure}[ht]
\includegraphics[scale=.7,center]{"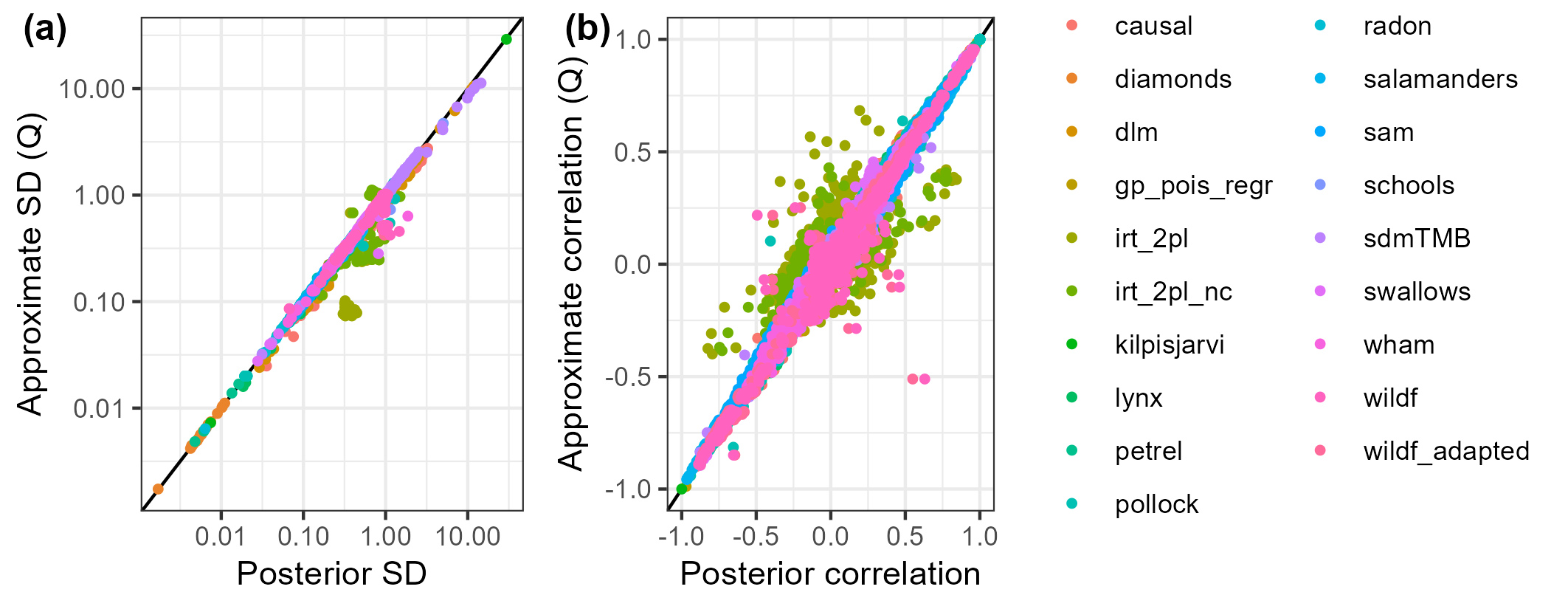"}
\caption{{\bfseries Evaluating $Q$ as a posterior.} The plots compare how well $Q$ can approximate the posterior distribution by model (colors) for (a) marginal standard deviations (SD) and (b) pairwise correlations. Points away from the 1:1 line suggest a multivariate normal with precision $Q$ is a poor approximation to the posterior.}
\label{fig:post_vs_Q}
\end{figure}

\begin{figure}[ht]
\includegraphics[scale=.7, center]{"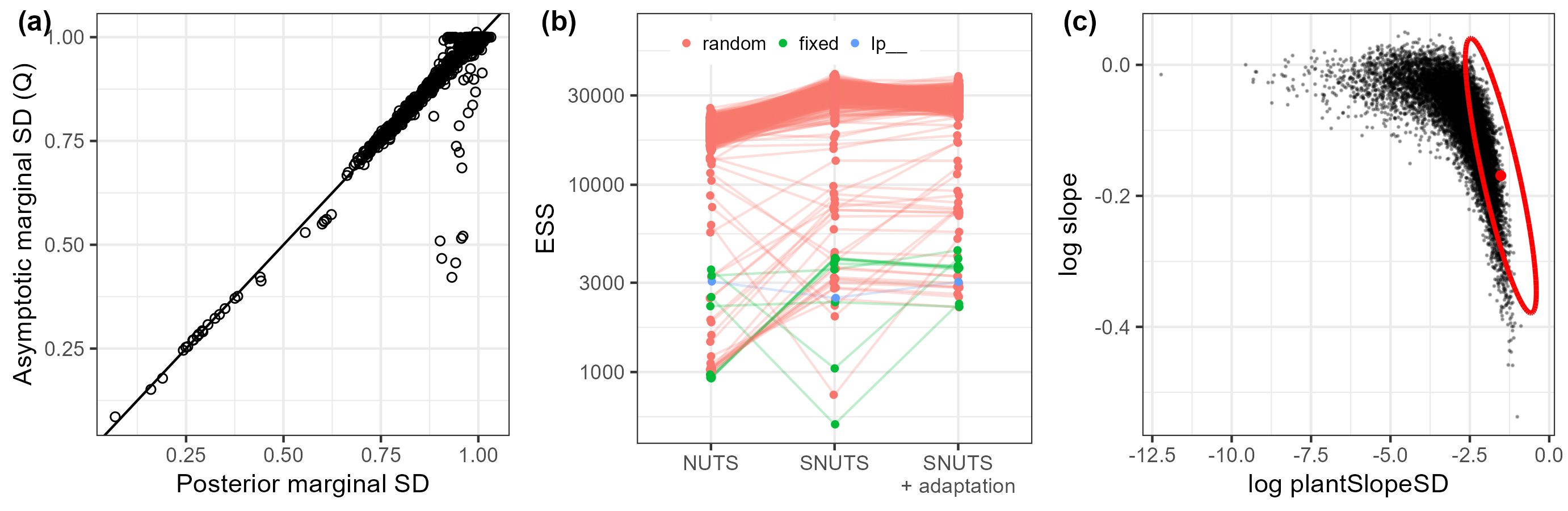"}
\caption{{\bfseries Investigating the poor performance of \model{wildf}}. The \model{wildf} model was rerun three times: standard NUTS (i.e., Stan defaults) which includes diagonal mass matrix adaptation during warmup, SNUTS defaults which skips mass matrix adaptation, and SNUTS again but with diagonal mass matrix adaptation turned on, with all three using four chains and \texttt{adapt\_delta=0.999} and 4000 total samples (1000 warmup) to get better convergence and stable estimates. (a) The marginal standard deviation for each parameter (point) from the posterior samples and the asymptotic approximation from the square root of the diagonal of  $Q^{-1}$ are compared for SNUTS. There is poor agreement in many of the parameters, indicating that $Q$ does not approximate the posterior well. (b) The effective sample size (ESS) for each parameter for the three runs (line segments) which are colored by whether the parameter is a fixed or random effect or is the unnormalized log posterior density (``lp\_\_''). A handful of parameters have a much lower ESS with SNUTS, but higher again with adaptation turned on, confirming that a poor approximation of $Q$ was detrimental to sampling efficiency, and that this can be recovered by adapting a diagonal mass matrix in Stan. (c) Scatter plot of posterior points for the two parameters with the lowest ESS, with the bivariate asymptotic normal 95\% ellipse from $Q$ (red lines) centered at the estimated mode. The approximation is poor for these two parameters.}
\label{fig:wildf_pairs}
\end{figure}

\begin{table}[ht]
\footnotesize
\centering
\begin{tabular}{p{2.5cm}p{10cm}c}
\hline
\textbf{R   package} &   \textbf{Title / Description} &    \textbf{Reference} \\
\hline
\package{aniMotum} & Fit latent variable movement models to animal tracking data for location quality control and behavioural inference. & \cite{jonsen2023animotum}\\
\package{ctsmTMB} &   Continuous-time stochastic modeling for state-space systems using the   Laplace approximation in \package{TMB}. &  \cite{vetter2025} \\
\package{dsem} &   Applies dynamic structural equation models to time-series data with   generic and simplified specification for simultaneous and lagged effects. &   \cite{thorson2024} \\
\package{ecostate} &   State-space mass-balance models for marine ecosystems. &   \cite{thorson2025} \\
\package{fishgrowth} &   Git growth curves to fish data. &   \cite{magnusson2025} \\
\package{gllvm} &   Analyzes multivariate ecological data using Generalized Linear Latent   Variable Models (GLLVM). &  \cite{niku2019} \\
\package{glmmTMB} &   Fits generalized linear mixed models (GLMMs) with various distributions   (Poisson, Negative Binomial, Zero-inflated) using Template Model   Builder. &   \cite{brooks2017} \\
\package{hmmTMB} &  A flexible framework for Hidden Markov Models with random effects and   smoothing splines. &  \cite{michelot2025} \\
\package{LaMa} &   Fast numerical maximum likelihood estimation for latent Markov models. &  \cite{mews2025} \\
\package{marked} &   Mark-recapture analysis for survival and abundance estimation. &   \cite{laake2013} \\
\package{mmrm} &   Mixed models for repeated measures (MMRM) are a popular choice for   analyzing longitudinal continuous outcomes in randomized clinical trials and   beyond. &  \cite{bove2025} \\
\package{phylosem} &  Implements phylogenetic structural equation models (SEM) combining   comparative methods and mixed models. &
  \cite{thorson2023phylosem} \\
\package{sdmTMB} &   Fast, flexible, and   user-friendly generalized linear mixed effects models with spatial and   spatiotemporal random fields. &  \cite{anderson2025} \\
\package{stochvolTMB} &   Parameter estimation for stochastic volatility models using maximum   likelihood. &   \cite{wahl2025} \\
\package{stockassessment} &   State-space assessment model used for fisheries stock assessment. &   \cite{nielsen2014} \\
\package{tinyVAST} &   Expressive interface to specify lagged and simultaneous effects in   multivariate spatio-temporal models. &   \citep{thorson2025b} \\
\package{tsissm} &   Unobserved components time series model using the linear innovations   state space representation (single source of error) with choice of error   distributions and option for dynamic variance. &  \cite{galanos2025} \\
\package{unmarked} &
  Fits hierarchical models for wildlife survey data (occupancy, abundance)   using \package{TMB} for fast marginal likelihood estimation. &
  \cite{kellner2023} \\
\package{VAST} &   Conducting spatio-temporal   analysis of data from multiple categories (species, sizes, etc.), and   includes both spatial and spatio-temporal variation in density for each   category, with either factor-analysis or autoregressive correlations among   categories, with derived calculation of abundance indices, center-of-gravity,   and area-occupied. &   \cite{thorson2019} \\
\package{wham} &   A general state-space assessment framework that incorporates time- and   age-varying processes via random effects and links to environmental   covariates. &   \cite{stock2021} \\
\package{tinyVAST} &   Expressive interface to specify lagged and simultaneous effects in   multivariate spatio-temporal models. &  \citep{thorson2025b}\\
  \hline
\end{tabular}
\vspace*{4pt}
\caption{{\bfseries Packages using \package{TMB}.} A non-exhaustive list of R packages that use Template Model Builder (TMB) or the R interface \package{RTMB} as a computational engine.}\label{tab:rpackages}
\end{table}

% \begin{landscape}
\begin{table}[ht]
\centering
\begin{tabular}{p{2cm}p{12cm}}
\hline
\footnotesize
\textbf{Name} & \textbf{Description and modifications}\\
\hline
\model{causal} &
 Multi-causal ecosystem synthesis   time-series model as implemented in ‘dsem’ \citep{thorson2024}. $\mathcal{N}(0,   0.25)$ priors were added to beta\_z[9:16] and $\mathcal{N}(0, 0.5)$ to beta\_z[20].   These parameters were weakly identified and had unrealistically long tails. \\
\model{diamonds} &
Linear mixed effects model of diamonds with correlated  covariates. Listed in \cite{magnusson2024} as one with a difficult geometry  for NUTS. \\
\model{dlm} &
Dynamic linear  model \url{https://james-thorson-noaa.github.io/dsem/articles/features.html#comparison-with-dynamic-linear-models}  implemented as in ‘dsem’  \citep{thorson2024}. \\
 \model{gp\_pois\_regr} &
 A Poisson Gaussian process regression. From the posteriordb  data base \citep{magnusson2024}. \\
 \model{irt\_2pl} &
 Item response theory model taken from the posteriordb data  base \citep{magnusson2024}. This is the centered version, and a non-centered  version irt\_2pl\_nc is shown in some figures. \\
 \model{kilpisjarvi} &
  Linear regression of annual summer temperatures with an uncentered covariate leading to high posterior correlations. Taken from  \cite{bales2019}. \\
\model{lynx} &
 Lynx and hare dynamics expressed as an ODE and solved with the fourth order Runge-Kutta method. Data were  retrieved from \url{https://jmahaffy.sdsu.edu/courses/f09/math636/lectures/lotka/qualde2.html}. \\
 \model{petrel} &
  A 2-state Hidden Markov model of movement with random effects for individual transition probabilities and gamma observations using hmmTMB \citep{michelot2025} fitted to Antarctic petrel data \citep{descamps2016}. \\
 \model{pollock} &
  A custom-built non-linear state-space fisheries population  dynamics model, adapted from \cite{monnahan2024}. We added stronger  priors for selectivity parameters for survey 2, and turned off estimation of  log\_DMpars[6] corresponding to the Dirichlet-multinomial overdispersion term  for survey 6 age compositions. \\
 \model{radon} &
  Linear mixed effects model of radon levels by country. Taken  from \cite{bales2019}. \\
 \model{salamanders} &
  A zero-inflated negative binomial model of salamander species  counts by site as fitted in glmmTMB \citep{brooks2017}. \\
\model{sam} &
Non-linear, state-space fisheries population dynamics fitted  in the ``sam'' package \citep{nielsen2014} and taken from the online examples at \url{https://github.com/fishfollower/SAM?tab=readme-ov-file#a-quick-example}. \\
\model{schools} &
 The classic non-centered version of 8-schools linear mixed  model of student scores as in e.g., \cite{gelman2014}. \\
\model{sdmTMB} &
 A spatial SPDE Tweedie GLMM with spline on depth covariate,  fit via the \package{sdmTMB} package \citep{anderson2025}. Modified from an example so that the parameter $\kappa$ was not estimated. \\
\model{swallows} &
 State-space survival and detection with environmental covariates fitted to mark-recapture data of birds, taken from  \cite{korner2015} and also examined in \cite{monnahan2017}. The normal prior for sigmayear was tightened to $\mathcal{N}(0,2)$. There were many NUTS divergences at  higher values for this parameter otherwise. \\
\model{wham} &
 A non-linear state-space stock assessment model using package  ``wham'' \citep{stock2021}, based on the example from the vignette with 2D AR(1) log numbers at  age states. Selectivity, and covariate regression parameters (Ecov) were not estimated due to exremely poor identifiability and lack of interface to assign priors instead.\\
\model{wildf} &
 Binomial GLMM of flowering success with year effects on  intercept and crossed effects on intercept and slope for a covariate, taken  from \cite{bolker2013} and also analyzed in depth in \cite{monnahan2017}. \\
\hline
\end{tabular}
\vspace*{6pt}
\caption{{\bfseries Case study details.} This table provides further details of the case studies used (see Table~1)}\label{tab:mods2}

\end{table}
% \end{landscape}

\begin{table}[ht]
\centering
\begin{tabular}{ccrrcc}
\hline
\textbf{Model} & \textbf{Metric} & \textbf{Parameters} & \textbf{\% sparsity of $Q$} & \textbf{Gradient ratio} & \textbf{\% for $Q$} \\
\hline
causal         & sparse & 264   & 78.7 & 1.19 & 0.2 \\
diamonds       & dense  & 26    & 0.0  & 1.00 & 2.2 \\
dlm            & dense  & 33    & 8.1  & 1.06 & 1.5 \\
gp\_pois\_regr & dense  & 13    & 0.0  & 1.10 & 0.0 \\
irt\_2pl       & sparse & 144   & 56.3 & 1.21 & 0.1 \\
irt\_2pl\_nc   & sparse & 144   & 56.2 & 1.24 & 0.1 \\
kilpisjarvi    & dense  & 3     & 0.0  & 1.12 & 0.1 \\
lynx           & dense  & 50    & 63.0 & 1.09 & 2.4 \\
petrel         & sparse & 274   & 94.3 & 1.00 & 1.1 \\
pollock        & sparse & 351   & 28.3 & 1.11 & 2.1 \\
radon          & diag   & 389   & 98.5 & 1.01 & 0.2 \\
salamanders    & dense  & 39    & 44.5 & 1.03 & 1.4 \\
sam            & sparse & 1226  & 96.6 & 1.43 & 1.2 \\
schools        & diag   & 10    & 62.2 & 1.08 & 1.0 \\
sdmTMB         & sparse & 219   & 81.3 & 1.01 & 0.0 \\
swallows       & sparse & 177   & 75.4 & 1.05 & 1.0 \\
wham           & sparse & 385   & 66.2 & 1.02 & 4.6 \\
wildf          & sparse & 1101  & 98.4 & 1.14 & 1.0 \\
\hline
\end{tabular}
\vspace*{8pt}
\caption{{\bfseries SNUTS Benchmarks.}  The table shows the computational costs of SNUTS on the case studies. The gradient ratio is the cost of $g_q(q)$ divided by $g_{q'}(q')$ which depends on the selected preconditioner, and represents how many times slower the transformed gradient is than the gradient. The last column the is computational cost of optimizing, calculating $Q$,  inverting it to get $Q^{-1}=\Sigma$, and doing Cholesky decompositions on $Q$ and $\Sigma$, as a percentage of the total SNUTS run time. See Table~1 for further model details.}\label{tab:timings}
\end{table}

\clearpage
\section{R source code}

This section contains relevant details of the R implementation.

\begin{Verbatim}
library(RTMB)
rm(list=ls())

# Set up 8 schools example:
dat <- list(J = 8,
            y = c(28,  8, -3,  7, -1,  1, 18, 12),
            sigma = c(15, 10, 16, 11,  9, 11, 10, 18))
pars <- list(eta=rep(1,8),mu=0, logtau=1)
f <- function(pars){
  RTMB::getAll(dat, pars)
  theta <- mu + exp(logtau) * eta;
  lp <- sum(dnorm(eta, 0,1, log=TRUE))+ # hyperprior
    sum(dnorm(y,theta,sigma,log=TRUE))+ # likelihood
    logtau                              # jacobian
  gq <- sum(exp(logtau)*eta[2:4])       # generated quantity fixed + random
  ADREPORT(gq)                          # request delta method SE
  REPORT(gq)                            # simple output
  return(-lp)                           # TMB uses negative lp
}
obj <- MakeADFun(func=f, parameters=pars, random='eta', silent=TRUE)

# Optimize and get Q via sdreport function
opt <- with(obj, nlminb(par,fn,gr))
sdrep <- sdreport(obj, getJointPrecision=TRUE)
str(sdrep$jointPrecision)
\end{Verbatim}

Then we manually calculate $Q$ using \package{TMB}'s AD machinery and the Matrix package functionality.

\begin{Verbatim}
# This code block shows how to calculate Q in R. It reproduces
# the 'sdreport' functionality in a minimal way (see equations
# 9-10) and only for RTMB models. It is meant only as a
# demonstration for the model above (schools).

library(Matrix)
q_hat <- obj$env$last.par.best # joint mode
n <- length(q_hat)
theta_hat <- opt$par          # marginal mode
r <- obj$env$random           # index of random effects
nonr <- setdiff(seq_along(q_hat), r)

# Hessian block for fixed effects using finite differences
H_Bhat <- optimHess(theta_hat, obj$fn, obj$gr) # Hessian of marginal posterior

# Hessian of random effects at joint mode using AD.
H_AA <- obj$env$spHess(q_hat, random = TRUE)

# Second derivatives of the joint posterior at the joint
# mode for the fixed:random effect elements only. Uses AD.
H_AB <- obj$env$f(q_hat, order = 1, type = "ADGrad", keepx=nonr, keepy=r) ## TMBad only !!!
H_BA <- t(H_AB)
H_BB <- H_BA %*% solve(t(H_AA)) %*% H_AB + H_Bhat
Q <- Matrix(0, nrow = n, ncol=n, sparse=TRUE)
Q[r,r] <- H_AA
Q[r,nonr] <- H_AB
Q[nonr,r] <- H_BA
Q[nonr, nonr] <- H_BB

# this matches TMB's internal calculations in sdreport:
max(abs((Q-sdrep$jointPrecision))) # [1] 2.775558e-17
\end{Verbatim}

Next we show how $Q$ can be used to decorrelate the parameter space (equation 3).

\begin{Verbatim}
# Use sparse metric Q to transform from original `x` to reparamterized `y`

# first rebuild object without the Laplace appoximation (random=NULL)
obj2 <- RTMB::MakeADFun(func=obj$env$data, parameters=obj$env$parList(),
                        map=obj$env$map,
                        random=NULL, silent=TRUE,
                        DLL=obj$env$DLL)
# Do Cholesky on permuted Q
chd <- Matrix::Cholesky(sdrep$jointPrecision, super=TRUE, perm=TRUE)
L <- as(chd, "sparseMatrix")
# Drop all numerical zeros and convert to triangular storage
L <- tril(drop0(L)) ## class(L) == "dtCMatrix"
Lt <- Matrix::t(L) ## class(Lt) == "dtCMatrix"
perm <- chd@perm + 1L            # the new order of the system
iperm <- Matrix::invPerm(perm)   # the inverse
# assume an initial value in x space, like the joint mode
x0 <- q_hat      # the joint mode
x0perm <- x0[perm]               # permuted
# initial value in the transformed, permuted space
y0perm <- as.vector(Lt %*% x0perm)
# redefine the objective and gradient functions
fn.y <- function(y)  -obj2$fn(Matrix::solve(Lt, y)[iperm])
# carefully implement this to optimize sparsity and reduce
# computations since the gradient call is the expensive part of
# NUTS
gr.y <- function(y){
  x <- Matrix::solve(Lt, y)[iperm]
  -Matrix::solve(L, as.numeric(obj2$gr(x))[perm])
}
# back transform parameters
y.to.x <- function(y) as.numeric(Matrix::solve(Lt, y)[iperm])

# test them out
fn.y(y0perm)
obj2$fn(x0)  # matches but for sign
gr.y(y0perm) # gradient at joint mode is 0 for fixed effects only

\end{Verbatim}

Finally, we implement SNUTS via the \package{StanEstimators} interface and show how to use the generalized delta method

\begin{Verbatim}
## -------------------------------------------------------------
# Now show how to link this through StanEstimators
library(StanEstimators)
fit <- stan_sample(fn=fn.y, par_inits=x0perm, grad_fun=gr.y,
                   num_chains=4, seed = 12345)
# posterior draws in transformed space
post.y <- unconstrain_draws(fit) |> as.data.frame()
# recorrelate the draws into untransformed space x
post.x <- t(apply(post.y[,2:11], 1, FUN=y.to.x))
cbind(postmean=colMeans(post.x), postmode=x0)

## -------------------------------------------------------------
# Now compare approximate (asymptotic normal) estimate of a
# generated quantity using the generalized delta method (via
# TMB::sdreport) against the posterior

# The mean and SE from the delta methd are assumed to be normal
gq.mle <- c(mean=sdrep$value, sd=sdrep$sd)
# push each posterior draw through and extract the generated
# quantity 'gq' to get posterior of gq
gq.post <- apply(post.x, 1, \(x) obj2$report(x)$gq)
hist(gq.post, freq=FALSE, breaks=30, main='Generated quantity example',
     xlab='gq')
x <- seq(min(gq.post), max(gq.post))
y <- dnorm(x, gq.mle[1], gq.mle[2])
lines(x,y, col=2, lwd=2)
\end{Verbatim}

\end{document}